\documentclass[prd,twocolumn,nofootinbib,notitlepage,aps,tightenlines,preprintnumbers,amsmath,amssymb,amsfonts,showpacs,superscriptaddress]{revtex4-2}

\input{header}

\begin{document}

\title{Tuning Pythia for Forward Physics Experiments}

\author{Max~Fieg}
\affiliation{Department of Physics and Astronomy, University of California, Irvine, CA 92697-4575, USA}

\author{Felix Kling}
\affiliation{Deutsches Elektronen-Synchrotron DESY, Notkestr.~85, 22607 Hamburg, Germany}

\author{Holger~Schulz}
\affiliation{Tharsus Ltd., Coniston Rd, Blyth NE24 4RF, United Kingdom}

\author{Torbj\"orn Sj\"ostrand}
\affiliation{Department of Physics, Lund University, Box 118, 221 00 Lund, Sweden}

\begin{abstract} 
Event generators like \pythia play an important role in physics studies at the Large Hadron Collider (LHC). While they make accurate predictions in the central region, \ie at pseudorapidities $\eta<5$, a disagreement between \pythia and measurements in the forward region, $\eta>7$, has been observed. We introduce a dedicated forward physics tune for the \pythia event generator to be used for forward physics studies at the LHC, which uses a more flexible modelling of beam remnant hadronization and is tuned to available particle spectra measured by LHCf. Furthermore, we provide an uncertainty estimate on the new tune in a data-driven way which can be used as a means of flux uncertainty for future forward physics studies. We demonstrate an application of our tune by showing the updated neutrino and dark photon spectra at the FASER experiment.
\end{abstract}
\maketitle


\section{Introduction}

The Large Hadron Collider (LHC) has been instrumental in constraining physics both within and beyond the Standard Model. Its main experiments, ATLAS, CMS, LHCb and ALICE, have discovered and measured properties of the Higgs, constrained dark sectors, probed new physics in the flavor sector, and more generally, have furthered our understanding of fundamental particle physics. These experiments benefit greatly from Monte Carlo event generators, which can make accurate predictions of particle distributions in the central region with pseudorapidities $\eta \lesssim 5$. Much work has been put into improving, validating and tuning these generators for the experiments at the LHC, and often excellent agreement has been reached. 

Recently, there has been new interest in particle production in the forward direction at the LHC, corresponding to $\eta \gtrsim 7$, where much less data has been collected as compared to the central experiments. The implementation of the FASER experiment has already set leading bounds in certain BSM scenarios~\cite{FASER:2023tle} and lead to the first direct observation of neutrinos produced at a collider~\cite{FASER:2021mtu, FASER:2023zcr}. Additionally, the Forward Physics Facility (FPF) has been proposed to house a suite of experiments to further study particles produced in the forward direction during the high-luminosity LHC era~\cite{Anchordoqui:2021ghd, Feng:2022inv}. The success of these experiments will be greatly enhanced if similar event generators can be used to make accurate predictions.

However, in the context of the LHC, the popular event generator \pythiafs~\cite{Sjostrand:2014zea, Bierlich:2022pfr} has only been tuned in the central region, and thus one should not expect reliable predictions in the forward direction. Indeed, the LHCf experiment, which can measure distributions of neutral particles with $\eta\gtrsim 9$, shows a distinct disagreement with \pythiafs's predictions obtained using the popular tune relying on data from central experiments --- the so-called \textit{Monash} tune~\cite{Skands:2014pea}. Notably, \pythia predicts an excess of mesons but a deficit of baryons when compared to LHCf data~\cite{LHCf:2015nel, LHCf:2020hjf, LHCf:2015rcj, LHCf:2017fnw}. 

In this paper we provide a forward physics tune for the \pythia event generator by fitting hadronization parameters to LHCf measurements of neutral pion, photon and neutron production. In particular, we will primarily fit parameters that have little impact on central physics, so as to not spoil the success of \pythia in this region. 

In addition to our forward tune, we will also provide an uncertainty estimate on these parameters. Currently, existing generators typically only provide one central prediction but no measure of uncertainty. One approach often used in astroparticle physics is to define an uncertainty based on the spread of event generators' predictions. While this definition captures a spread of underlying physics modelling, it is not data-driven and it is not clear if it has any statistical meaning. Here, for the first time, we follow a different approach and provide the uncertainty on a single generator in a data-driven way.

This paper is organized as follows. In \cref{sec:PythiaModelling}, we discuss how hadronization is done in \pythia in the forward direction. In \cref{sec:tuningkinematics} we discuss our tuning procedure to the LHCf measurements and provide our tune on these kinematic parameters. In \cref{sec:ApplicationAtFPF}, we show how our tune impacts the predictions for forward neutrino and dark photon production at the FASER experiment. In \cref{sec:Conclusion}, we summarize and conclude.

\section{Modeling of Forward Particle Production in Pythia}\label{sec:PythiaModelling}

There are few theory constraints in the modelling of forward physics. 
While at least some aspects of central physics are governed by 
perturbation theory, such as jet production, the forward region
is entirely of nonperturbative origin. 

An early assumption was so-called Feynman scaling~\cite{Feynman:1969ej},
\ie\ that the $x_E \, \mathrm{d}n /\mathrm{d}x_{\mathrm{F}}$ distribution
should be collision-energy-independent. Here 
$x_{\mathrm{F}} =  2 p_z / E_{\mathrm{CM}}$ and
$x_{\mathrm{E}} =  2 E / E_{\mathrm{CM}}$
in the rest frame of the event, and $n$ is the number of produced particles
per event. Perfect Feynman scaling would correspond to a
collision-energy-independent central rapidity plateau
$\mathrm{d}n /\mathrm{d}y$, while data instead show this distribution
to be rising with energy, suggesting that an increasing fraction of the
total energy is taken from the forward region to produce more particles
in the central one. 

Central particle production in \pythia is generated by multiparton 
interactions (MPIs). That is, since hadrons are composite objects, 
several parton--parton subcollisions can occur inside a single $pp$ 
event. The high-$\pT$ tail of these subcollisions corresponds to regular
jet production. The bulk of them occur at a few GeV, however, where they
are not distinguished individually but are only visible by their
collective effect. The rise of the rapidity plateau is mainly driven
by an increasing average number of MPIs.

The beam remnants stem from those partons that are \textit{not} kicked 
out of the incoming protons by MPIs. The remnants and the MPIs are related 
to each other by flavor and color. An MPI can take away both valence
and sea quarks from the original $uud$ valence content of a proton,
giving rise to varied remnant topologies, \eg\ $ud$ or $uud\overline{s}$.
Each kicked-out (anti)quark also carries away some (anti)color, and
each gluon both a color and an anticolor, that have to be
compensated in the remnant so as to preserve the overall color
state. In the Lund string model~\cite{Andersson:1983ia}, each separated
color--anticolor pair gives rise to a linear confinement field,
a string, that will fragment into hadrons. This would mean that 
the momentum of a remnant often had to be shared between many 
string systems, making it difficult to obtain a leading baryon that
carries a significant fraction of the incoming proton energy. 
Also note that the number of MPIs goes up with increasing collision 
energy, implying softening baryon spectra.

Indeed, the problem in \pythia is to produce a spectrum with a fair 
amount of high-momentum baryons, and some corrections have to be 
introduced to the baseline picture, as will be outlined in this section.
We here do not consider the class of elastic scattering, which
obviously is quite separate and not of interest here. We also leave
diffraction aside for now but return to it later.   
 
Early on~\cite{Sjostrand:1987su} it was realized that a picture of fully
independent MPIs does not reproduce collider phenomenology, \eg\ the
rise of the average transverse momentum of charged particles with
increasing multiplicity. Hence the need for color reconnection (CR),
the assumption that nature has a tendency to rearrange colors
such that the total string length to be drawn out is reduced.
Many possible scenarios have been proposed over the years, and a few
of them are implemented in \pythiafs. We will here study two of them.

In the default CR scenario, it is assumed that the partons pulled out
from the colliding protons are strongly correlated in color, in a way
that the color of one such parton may be the same as the anticolor of
another such. In a picture where only gluons are pulled out, the resulting
remnant would then be in a color octet state, which conveniently can
be subdivided into a triplet single quark and an antitriplet diquark.  
If in addition one valence quark is kicked out, only a diquark remains.
These are the two most common outcomes, but others are possible and
modelled. One is that all three valence quarks are kicked out. Then
a single gluon is assigned to carry the remaining energy and momentum.
Another is that the removal of sea quarks leaves their antipartners
behind. Then the remnant is simplified by splitting off a hadron, \eg\
$uud\overline{s} \to ud + u\overline{s} \to ud + K^+$.

The other scenario is the QCDCR one~\cite{Christiansen:2015yqa}.
In it, explicit colors are assigned both to quarks and gluons, and
reconnections can occur between identical colors if they reduce the
total string length. Such a detailed tracing of color is not done in
the default scenario. Another distinguishing feature of QCDCR is
so-called junction reconnections. In it, two triplet strings can combine
into an antitriplet one, according to the familiar color algebra
$\mathbf{ 3 \otimes 3 = \overline{3} \oplus 6}$.
This leads to Y-shaped color topologies that carry non-vanishing
baryon numbers. Notably, the QCDCR model correctly predicts an increased
fraction of charm baryons in $pp$ relative to $e^+e^-$ events, which
the default does not~\cite{CMS:2019uws, ALICE:2020wla}.

Zooming in on the remnant region, the QCDCR starting point is again
to assign explicit colors to each parton pulled out of the incoming
protons, with opposite colors in the remnant. This allows a
bigger color charge to accumulate in the remnant than assumed in
the default scenario, and this requires additional remnant gluons.
In a first instance the remnant is only simplified when \eg\
the color of one gluon equals the anticolor of another gluon. 
But again, high remnant color charges are deemed less likely,
so an exponential suppression in the size of the remnant multiplet
is introduced, whereby more remnant color lines are forced to 
cancel.

In the following, we will introduce a new forward physics tune that uses the QCDCR
scenario with its suggested parameter values~\cite{Christiansen:2015yqa} 
as a starting point. On top of that, some old or new parameters are 
varied, with a special eye towards consequences in the forward region.
An alternative tune that uses the default CR scenario and
the Monash tune~\cite{Skands:2014pea} as starting point, is presented
in \cref{sec:monashdiscussion}.

Whenever the remnant consists of more than one parton, the remnant
energy and (longitudinal) momentum have to be shared between them.
To this end, there are assumed shapes for valence and sea quark
momentum fractions $x$, as well as for gluons. With each $x$ first
picked at random according to these shapes, and then rescaled to
unit sum, each parton energy is now assigned a fraction
$x_{\textrm{rescaled}}$ of the full remnant energy. A diquark
receives the sum of the constituent quark $x$ values, but is in
addition allowed a further enhancement factor, by default 2. A
remnant hadron receives the sum of its constituent momenta. The
bottom line is that, in the two most common cases, either a diquark 
carries the full remnant momentum, or it carries an average of 
80\% of it.

It is this diquark that primarily can fragment to produce the leading
baryon, \eg\ the neutron measured by LHCf. In spite of the steps
already taken to make the diquark hard, it still turns out that the
default fragmentation results in too soft neutrons. We have therefore 
sought ways to further harden the leading baryon spectrum. This
requires modifications to the fragmentation of a leading diquark,
relative to the normal string scenario. 

To give some background, consider the normal string fragmentation, 
as probed most directly in $e^+e^-$ annihilation events, 
$e^+e^- \to \gamma^*/Z^0 \to q_0\qbar_0$. There the string 
between the $q_0$ and $\qbar_0$ breaks by the production 
of new $q_i\qbar_i$ pairs, to give a sequence 
$q_0\qbar_1 - q_1\qbar_2 - q_2\qbar_3 - \cdots - q_{n-1}\qbar_0$
of $n$ mesons. Here $q_0\qbar_1$ is called the first-rank hadron
of the $q_0$ jet, $q_1\qbar_2$ the second-rank one, and so on.
The simplest extension to baryon production is to allow also
antidiquark--diquark breaks, where the color antitriplet diquark
takes the role of an antiquark, and vice versa. Thereby the baryon
and antibaryon are nearest neighbors in rank, giving rise both to
flavor and momentum correlations. Specifically, since two flavor
pairs are shared, you could not produce a $\Xi - \overline{p}$ 
combination this way. Studies mainly at LEP have shown that
baryon--antibaryon pairs are more decorrelated than this picture
allows for.

This is where the popcorn mechanism enters. In it, diquarks are not bound objects, 
but quarks can drift apart along the string, in such a way that 
a meson can be produced between the baryon and antibaryon, whereby
the latter two only share one $q_i\qbar_i$ pair. Tunes to LEP 
data suggest that half the time the baryon and antibaryon are nearest   
neighbors, and half the time they are separated by a meson in between.
Translated to the fragmentation of a leading diquark, this means
that the production of a baryon and of a meson as the first-rank 
particle are about equally likely. But we do not have quite as nice 
a test bed for diquark fragmentation as $e^+e^-$ offers for quark one, 
and also have not spent a corresponding effort at tuning, so this 
assumption is untested. On the contrary, it is plausible that an
initial diquark from an incoming proton sticks together better than
assumed for new string breaks. Therefore, we introduce a new parameter,
$d_{\mathrm{pop}}$ (see \cref{table:parms} for the full name in the
code) uniquely for diquarks at the end of strings. If zero, then such 
a diquark will never break up, while if unity such a split is as 
likely as inside a string. A second-rank baryon takes less average 
momentum than a first-rank one does, so a reduced admixture of the 
former gives a harder baryon spectrum. 

For an initial parton in a string aligned along the $z$ axis, the 
first-rank hadron takes a fraction $z_1$ of the total lightcone 
momentum $E + p_z$, the second-rank a fraction $z_2$ of what is left 
after the first, \ie\ a fraction $z_2(1-z_1)$ of the original 
amount, and so on. In each step we assume the $z$ value to be picked 
at random according to the Lund symmetric fragmentation function 
(LSFF). In its full generality the LSFF allows for one separate 
parameter for each quark/diquark flavor species, and quark/diquark 
mass correction factors for the first-rank hadron. In practice this 
full generality is seldom used, and then the LSFF simplifies to
\begin{equation}
f(z) \propto \frac{1}{z} \, (1 - z)^a \, 
\exp \left( - \frac{b m_{\perp}^2}{z} \right) ~.
\label{eq:fz}
\end{equation}
Here $m_{\perp}^2 = m^2 + \pT^2$ is the squared transverse mass 
of the produced hadron, and $a$ and $b$ are free parameters to be
tuned. A relevant aspect is that hadrons with a larger mass also
take a larger average $z$ value. Nevertheless, it appears that the
forward baryon spectrum needs to be harder than is default.  
For the purposes of this tune we have therefore allowed $a$ and $b$ 
to be set separately when a diquark jet produces a first-rank baryon; 
hence the new parameters $a_{\mathrm{remn}}$  and $b_{\mathrm{remn}}$ which can be turned on by setting $f_{\mathrm{remn}} =$~on.
In a future, with more data and understanding at hand, alternative 
modifications could be considered.

In addition to the flavor and longitudinal structure of particle
production, also the transverse fragmentation must be considered.
Here the discussion can be split into the partonic setup and the 
string fragmentation.  

In the first stage, each parton taken out of the incoming proton to 
become part of an MPI is assumed to have some transverse motion, 
``primordial $\kT$''. This is expected to be of the order of the 
quark constituent mass, say a third of a GeV. For hard
processes, notably $Z$-boson production, empirically a higher scale
of order 2~GeV is required. This could be owing to an imperfect
modelling of the low-$\pT$ behavior of initial-state parton
showers, but whatever the reason an interpolation is introduced 
wherein soft systems receive a lower primordial $\kT$ and 
hard systems a higher one. The full expression for the Gaussian  
width $\sigma$ is
\begin{equation}
\sigma= \frac{\sigma_{\mathrm{soft}} \, Q_{\mathrm{half}} 
\!+\! \sigma_{\mathrm{hard}} \, Q}{Q_{\mathrm{half}} + Q}  
\frac{m}{m \!+\! m_{\mathrm{half}} \sqrt{\frac{E}{m}}} \ . 
\label{eq:sigma}
\end{equation} 
Here the $Q$, $m$ and $E$ are the hard scale, mass and energy of the 
MPI subsystem, while $\sigma_{\mathrm{soft}}$, $\sigma_{\mathrm{hard}}$, 
$Q_{\mathrm{half}}$ and $m_{\mathrm{half}}$ are free parameters.
The second factor is intended to reduce $\sigma$ for low-mass 
systems, especially if these are strongly boosted in the forward
direction ($E \gg m$). 

Also, the left-behind constituents of the beam remnants, mainly quarks
and diquarks, are each assigned a primordial $\kT$ with a
Gaussian width $\sigma_{\mathrm{remn}}$. Taken together, the MPI 
initiators and the remnant constituents add to give a net $\pT$. 
An opposite recoil is shared evenly by them all, except that the 
damping factor for low-mass systems in \cref{eq:sigma} is used
also here, such that transverse momentum overall is conserved.  
 
With the kinematics of partons fully defined, string fragmentation 
can be applied. Again consider a string piece aligned along the $z$ 
axis. Then, in each string break, the new $q_i$ and $\qbar_i$ are 
assumed to receive opposite and compensating $\pT$ kicks, which add 
vectorially to give the total $\pT$ of each $q_i\qbar_{i+1}$ hadron. 
Again a Gaussian distribution is used, with width $\sigma$. The full 
$\pT$ of a hadron is then obtained after the rotation and boost back 
to the full $pp$ collision frame, which notably depends on the 
primordial $\kT$ assigned to the remnant in the previous step.

A final note. So far we have considered nondiffractive events.
Diffraction in \pythia is based on the Ingelman--Schlein picture
\cite{Ingelman:1984ns}, wherein a diffractive system can be modelled
as a proton--glueball collision, where the glueball ``hadron'' is viewed 
as a representation of a pomeron. Notably the proton end of this system,
which is in the forward direction, is next-to identical with the one
of a nondiffractive system. The glueball end usually is at more central 
rapidities, and has negligible impact on the forward region.
The picture is slightly modified for low-mass diffraction, but
is there assumed dominated by the production of a string with one
leading diquark. Therefore, the modifications already introduced for
nondiffractive events can be reused, without the introduction of any
further ones.

In summary, the two main new modifications of the \pythia code are 
to allow a reduced probability for a remnant diquark to break up,
and to allow a harder fragmentation function for it. In addition,
some existing parameters are also modified within the tuning effort.

\begin{table*}[t]
\begin{tabular}{l|c||c|c|c}
\hline\hline
 Full name & Shorthand & Baseline (QCDCR) & Forward Tune & Uncertainty\\ \hline
 \texttt{BeamRemnants:dampPopcorn} & $d_{\mathrm{pop}}$ & 1 & 0 &\\ 
 \texttt{BeamRemnants:hardRemnantBaryon} & $f_{\mathrm{remn}}$ & off & on & \\
 \texttt{BeamRemnants:aRemnantBaryon} & $a_{\mathrm{remn}}$ & - & 0.36 & \\
 \texttt{BeamRemnants:bRemnantBaryon} & $b_{\mathrm{remn}}$ & - & 1.69 &\\
 \texttt{BeamRemnants:primordialKTsoft} & $\sigma_{\mathrm{soft}}$ & 0.9 & 0.58 & $0.26 \dots 1.27$ \\
 \texttt{BeamRemnants:primordialKThard} & $\sigma_{\mathrm{hard}}$ & 1.8 & 1.8& \\
 \texttt{BeamRemnants:halfScaleForKT} & $Q_{\mathrm{half}}$ & 1.5 & 10 &\\
 \texttt{BeamRemnants:halfMassForKT} & $m_{\mathrm{half}}$ & 1 & 1 &\\
 \texttt{BeamRemnants:primordialKTremnant} & $\sigma_{\mathrm{remn}}$ & 0.4 & 0.58 & $0.26 \dots 1.27$ \\
 \hline \hline
\end{tabular}
\caption{The main \pythia parameters studied in this article, their default parameters in the QCDCR tune (according to the \textit{Mode~2} configuration in Ref.~\cite{Christiansen:2015yqa}), and their values in the Forward Physics Tune obtained in this study. The last column shows the uncertainty range for $\sigma_{\mathrm{soft}} = \sigma_{\mathrm{remn}}$ as discussed in \cref{sec:uncertainty}.}
\label{table:parms}
\end{table*}

\section{Tuning Kinematics}
\label{sec:tuningkinematics}

As described in the previous section, the modeling of forward particle production introduces a number of phenomenological parameters. Their role is to parameterize the inability to make first principle predictions in the absence of perturbative methods. For the simulation to have predictive power, it is imperative that these parameters are set to values ("tuned") in such a way that the simulation reproduces a wide range of measured datasets, in this case from LHCf. In this section, we first discuss the datasets, parameters and methodology before presenting the results in the form of a forward physics tune that is based on the QCDCR scenario. The tuning parameters and their values for both the baseline tune and the forward physics tune are shown in \cref{table:parms}. Results for an alternative tune that is based on the default CR scenario and the Monash tune are presented for comparison in \cref{sec:monashdiscussion}.

\subsection{Datasets}
\label{sec:datasets}

We exclusively use data measured by the LHCf experiment for tuning purposes in this study as it is by far the most relevant source of information on forward particle production. LHCf measured neutral hadron and photon fluxes at forward rapidities $\eta \gtrsim 8.8$~\cite{LHCf:2006kzv}.  It is worth noting that forward photon production is dominated by $\pi^0 \rightarrow \gamma \gamma$ decay. We reasonably assume that the same mechanisms govern hadronization mechanisms at $\sqrt{s}=$7~TeV and 13~TeV collision energies. We therefore use LHCf data from both energies. The following list is a summary of the LHCf datasets we use to tune our phenomenological parameters with:
\begin{itemize}
    \setlength\itemsep{-1mm}
    \item neutron energy spectra at 7~TeV~\cite{LHCf:2015nel}
    \item neutron energy spectra at 13~TeV~\cite{LHCf:2020hjf}
    \item $\pi^0$ energy spectra at 7~TeV~\cite{LHCf:2015rcj}
    \item photon $p_z$ spectra at 13~TeV~\cite{LHCf:2017fnw}
\end{itemize}
The data are publicly available in the form of histograms of cross-sections that are differential in either $\eta$ or $p_{\perp}$. 

We note that we use a very recently published LHCf measurement on $\eta$ mesons~\cite{Adriani:2023tyb} for validation of our methodology. We further validate our result by confronting the tuned simulation with more central measurements from CMS and TOTEM in \cref{sec:discussresults}.

\subsection{Tuning Parameters}
\label{sec:tuningparameter}
 
Our mission is to identify and tune the value of phenomenological parameters relevant to forward physics while at the same time keeping the excellent predictive power of \pythia for central physics intact. In this context, working with parameters related to the modeling of the beam remnants (\cref{table:parms}) is a natural choice. They predominantly influence forward particle production while, as we will show, their influence on central particle production is limited. In the following, we discuss the effects these parameters have on the predictions of forward particle spectra, how the parameters are tuned to data, and finally, we present a robust uncertainty estimate for the most relevant parameters.

Compared to the experimental data, the default \pythia configuration predicts too many hard pions in the LHCf phase-space. Disabling the popcorn mechanism for meson production from beam remnants (\ie setting ~$d_{\rm pop} = 0$) leads to the desired reduction of hard pions. We note that we studied the effect of varying $d_{\rm pop}$ but found only little sensitivity for small $d_{\rm pop}>0$ and hence set this parameter to $0$. A side-effect of disabling the popcorn mechanism in beam remnants is an increase in the production of hard neutrons, simply because remnant diquarks can no longer hadronize into mesons. This turns out to be fortuitous, as \pythia's default predicts too few hard neutrons in the most forward direction $\eta > 10.76$. 

By adjusting other parameters associated with the beam remnant, we can tune the overall normalization of the forward hadronic flux. In particular, we can modify the initial $k_{\perp}$ of the partons in the incoming protons: partons with a relatively larger $k_{\perp}$ will generally pull hadrons towards distributions of smaller $\eta$. The phenomenology of this effect is governed by the width of the primordial $k_{\perp}$ distribution for the MPI initiators. The corresponding tuning parameters are $\sigma_{\rm soft},\sigma_{\rm hard},~{\rm and } ~Q_{\rm half}$, and for the beam remnant, $\sigma_{\rm remn}$. The net effect is a non-zero $p_{\perp}$ imparted on hadrons, the manifestation of which can be seen in the forward neutron and pion spectrum. 

The overall effects of $\sigma_{\rm soft}, \sigma_{\rm hard}$ and $\sigma_{\rm remn}$ on Pythia's predictions for LHCf measurements are qualitatively similar while their sensitivities are not (See our discussion in \cref{sec:PythiaModelling}). An increase in any of these parameters makes it more likely that forward hadrons inherit larger transverse momenta and therefore populate more central phase-space regions (i.e. bins with smaller $\eta$ in the LHCf data). We exploit this freedom the model gives us and take a pragmatic approach. To keep $\sigma_{\rm hard}$ at its default value of 1.8 GeV, we reduce its sensitivity by increasing the (poorly constrained) $Q_{\rm half}$ to 10 GeV. As can be seen in \cref{eq:sigma}, this makes the $k_{\rm \perp}$ distribution more dependent on $\sigma_{\rm soft}$. To remove the remaining degeneracy between $\sigma_{\rm soft}$ and $\sigma_{\rm remn}$, which have default values of 0.9 GeV and 0.4 GeV, we define a parameter $\sigma$ that relates the two: $\sigma =\sigma_{\rm soft} = f~\sigma_{\rm remn}$, where $f$ is a number that fixes the ratio. We studied the effect of tuning $\sigma$ when choosing different values of $f$ in the vicinity of $f=1$. Since we found only marginal improvement, we choose to fix $f$ at a value of $f=1$ and keep only $\sigma$ as a tuning parameter.

Two parameters, $a_{\rm remn}$ and $b_{\rm remn}$, that govern the baryon fragmentation function complete our set of tuning parameters. They allow us to have an almost exclusive handle on the neutron spectrum, without much impact on the pion spectrum. In our setup, lowering (raising) $a_{\rm remn}$ while raising (lowering) $b_{\rm remn}$ results in slightly harder (softer) forward neutron spectra. Initially, we studied the effect of treating $a_{\rm remn}$ and $b_{\rm remn}$ as independent tuning parameters. However, we found that equally good quality of \pythia predictions can be achieved by fixing $a_{\rm remn}$ to the base tune's value for the LSFF of 0.36 and tuning only $b_{\rm remn}$.

\subsection{Tuning Methods}

The observations detailed in the previous paragraph lead to a reduction of the dimensionality of the tuning problem. We are left with two free parameters, $\sigma$ and $b_{\rm remn}$ which we will explore in the ranges of $\left[0 - 1.5\right]$ and $\left[0.5-5\right]$, respectively. We use \pythia to simulate 7 million $pp$ collisions at $\sqrt{s}=7$~TeV and 5 million collisions at 13 TeV for each point we initially explore in the ($\sigma$, $b_{\rm remn}$) space. We analyze the simulated events with \rivet, enabling the analysis routines that correspond to the experimental data listed in \cref{sec:datasets}. The result of the analyses is a set of histograms obtained from simulation that can immediately be compared with the corresponding experimentally measured histograms. It should be noted that we obtain a set of histograms for each point in the so-explored parameter space.

Equipped with experimentally measured histograms and a method to obtain simulated histograms for any point in the parameter space, we could define a goodness-of-fit measure and numerically find a best-fit point that minimizes the measure. However, the computational cost to do so is prohibitively expensive. Instead, we construct an analytic surrogate model of the simulation response to shifts in the parameter space. The model allows us to predict the simulation outcome at any point in the parameter space at a fraction of the cost of computing the actual simulation. Not only is the model cheap to evaluate but, due to its analytic nature, it is also straightforward to compute first- and second-order derivatives.  These qualities make it an ideal fit for numerical minimization. We use the \textsc{Apprentice} toolkit for event generator tuning~\cite{Krishnamoorthy:2021nwv} to facilitate the construction of the surrogate, the definition of a goodness-of-fit measure, and the minimization thereof.  We explored different options for the surrogates and found no benefit in going beyond quadratic polynomials. As input to the surrogate, we use the full simulation results at 64 uniformly distributed points in the specified range for $\sigma$ and $b_{\rm remn}$.

The \textsc{Apprentice} toolkit allows to bias the goodness-of-fit measure using a weighting mechanism for individual histograms and even bins. In general, one might wish to better reproduce either the neutron spectra, photon spectra, pion spectra, or a subset of certain $\eta$ bins. We, however, wish to be agnostic and place the neutron, photon, and pion spectra measured at LHCf on equal footing. Since the datasets under consideration have quite different numbers of bins, we decided on a democratic weighting such that each of the four analyses is normalized according to the number of data points in that analysis. For a given particle spectrum and collision energy from \cref{sec:datasets}, the weighting can be expressed as $w = (N_{\rm bins})^{-1}$ where $N_{\rm bins}$ is the number of data points across $\eta$ (or $p_{\perp}$) bins in that set. 

\textsc{Apprentice} is then used to minimize the weighted goodness-of-fit measure. The outputs are a best-fit point $\sigma_0, b_{{\rm remn},0}$, and predicted spectra at that point, computed from the surrogate model. These spectra are compared against the actual output of the simulation when run with the parameters of the best-fit point in a necessary effort to validate the method. The best-fit values for $\sigma$ and $b_{\rm remn}$ for our forward physics tune can be found in \cref{table:parms}. 

\subsection{Tuning Uncertainty}
\label{sec:uncertainty}

In addition to the central tuning prediction, we wish to provide a measure of uncertainty on our best fit. An approach to estimate the uncertainty sometimes used in astroparticle physics is taken to be the spread in different event generators' predictions. While this does capture differences in underlying physics modeling, this definition is not data-driven and the error band lacks statistical meaning. 

Naively, one might follow the usual method of looking for $\Delta \chi^2 = 1$ to obtain a 68\% confidence interval. However, due to unknown correlations in experimental data, and imperfections in the physics modeling, the goodness-of-fit measure does not follow a $\chi^2$ distribution.
If one were to nonetheless follow that approach with our model, the observed $\chi^2_{\rm min}$ results in an unusable underestimate of uncertainties.

In light of this, we take a more practical approach. Our goal is to provide a well-defined range for our tuning parameters that can return a spread of particle fluxes for future studies at the FPF. This range can be obtained by varying the prediction in the vicinity of the best-fit and testing how much the predictions change. The question remains: how much should one vary the tuning parameters to find the corresponding upper and lower bound? A practical parameter uncertainty range is one that covers distances of \pythia's prediction at the best-fit point from the experimentally measured data and data uncertainties.

We find that our fitting parameters, $\sigma$ and $b_{\rm remn}$, are not strongly correlated and that deviations about the best-fit point are most sensitive to $\sigma$. We therefore choose to vary and provide an uncertainty on $\sigma$. To obtain this uncertainty, we define a prediction band specified by two points, $(f\times \sigma_0 , \sigma_0 / f)$, where $f$ is a number that is increased until the band contains 68\% of points (for $f=1$ the band obviously contains zero points). Now, even for extremal values of $\sigma$ in our range, there are a small number of data points which \pythia has difficulty describing; the central value of these points lies just outside the prediction range specified by $\sigma\in[0 - 1.5]$ and are typically found in the highest or lowest bins of the energy spectrum. Since we do not want those points to drive our estimation of uncertainty, we exclude them when counting the fraction of points inside the band specified by $f$. Across the four analyses there are 20 of these out of 306 total data points.

The method yields two parameter points $\sigma_-,\sigma_+$ which define a robust uncertainty band containing 68\% of points: $0.26 < \sigma < 1.27$.

\subsection{Discussion of Results}
\label{sec:discussresults}


\begin{figure*}[tbh]
  \includegraphics[width=0.48\textwidth]{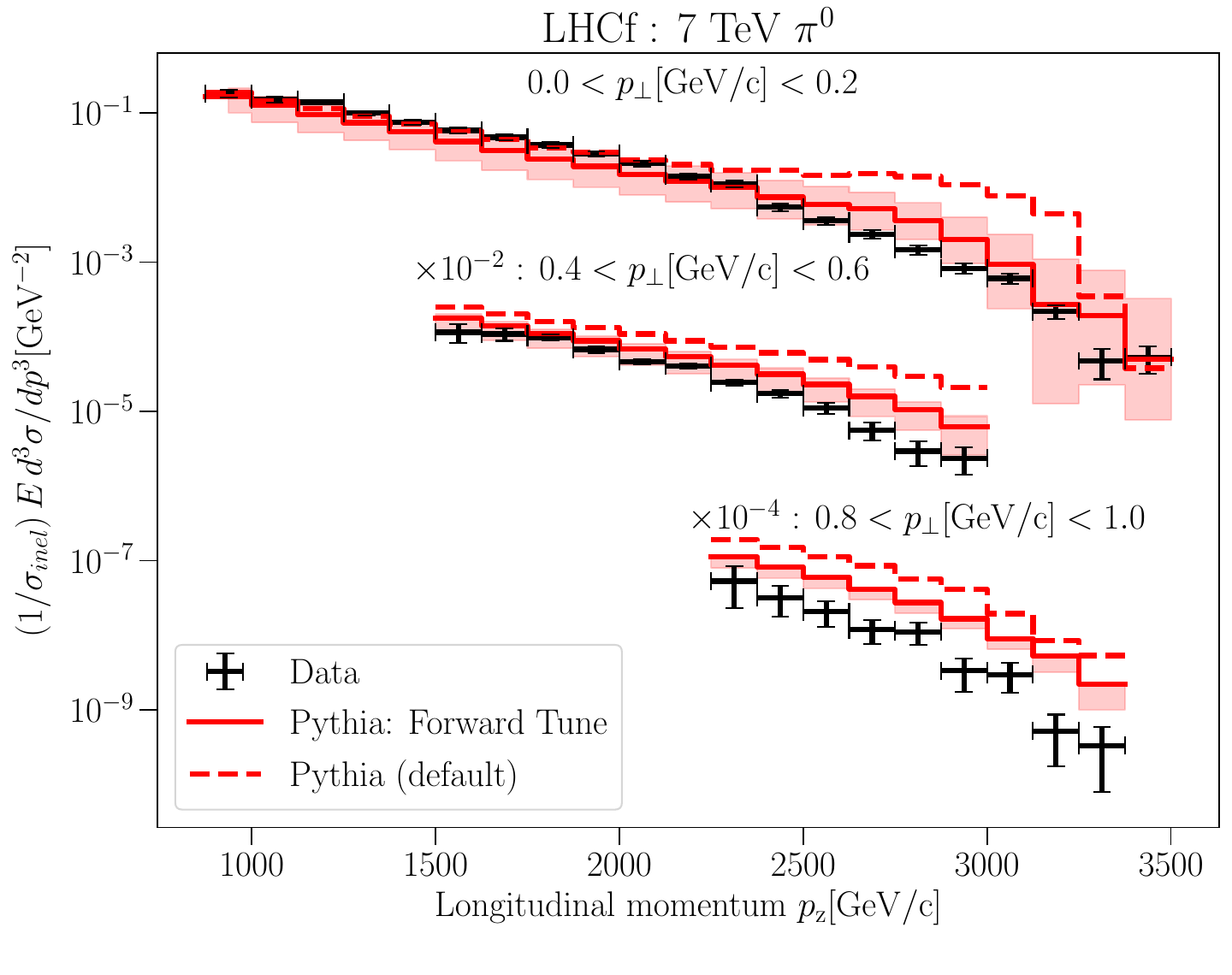}
  \includegraphics[width=0.48\textwidth]{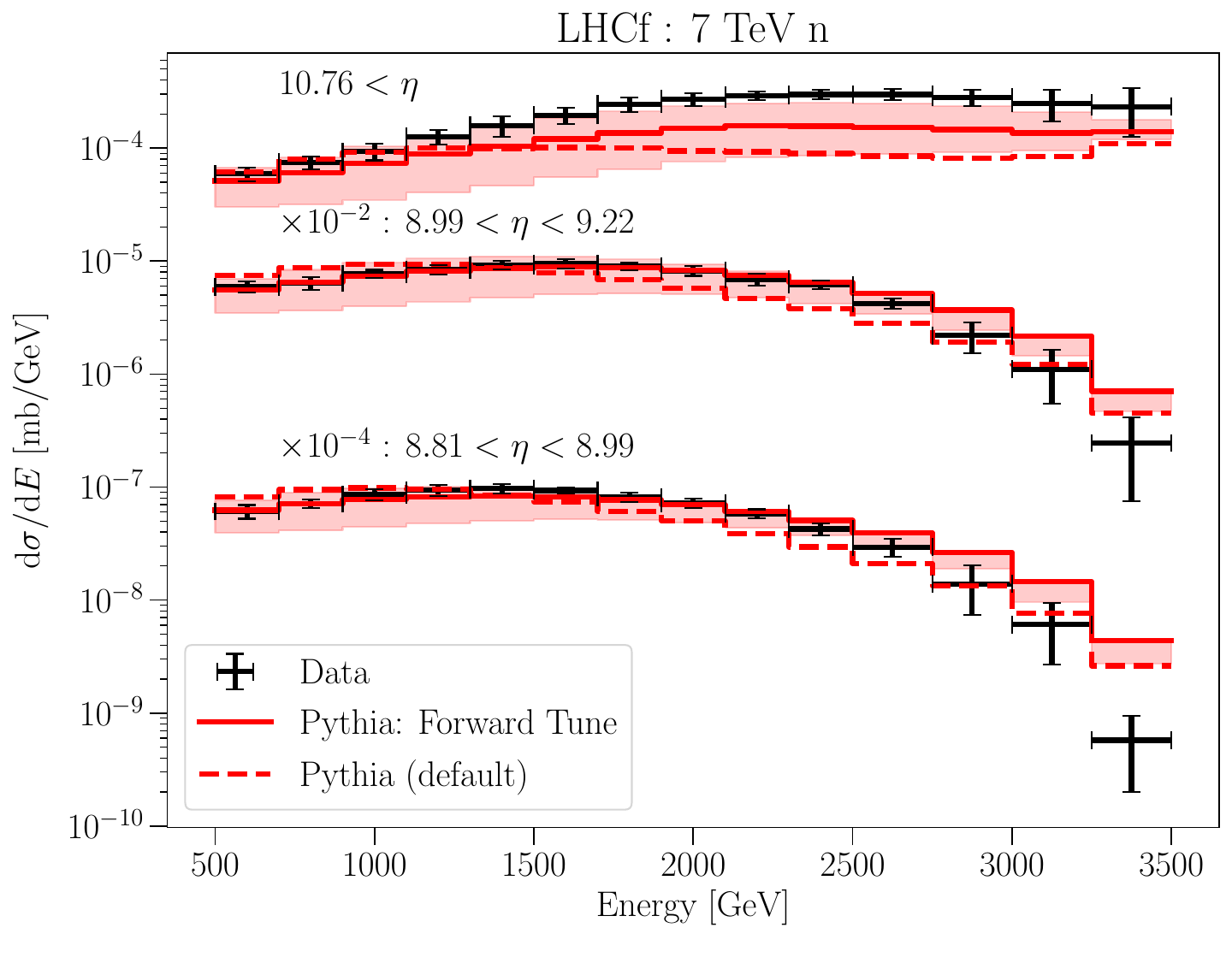}
  \includegraphics[width=0.48\textwidth]{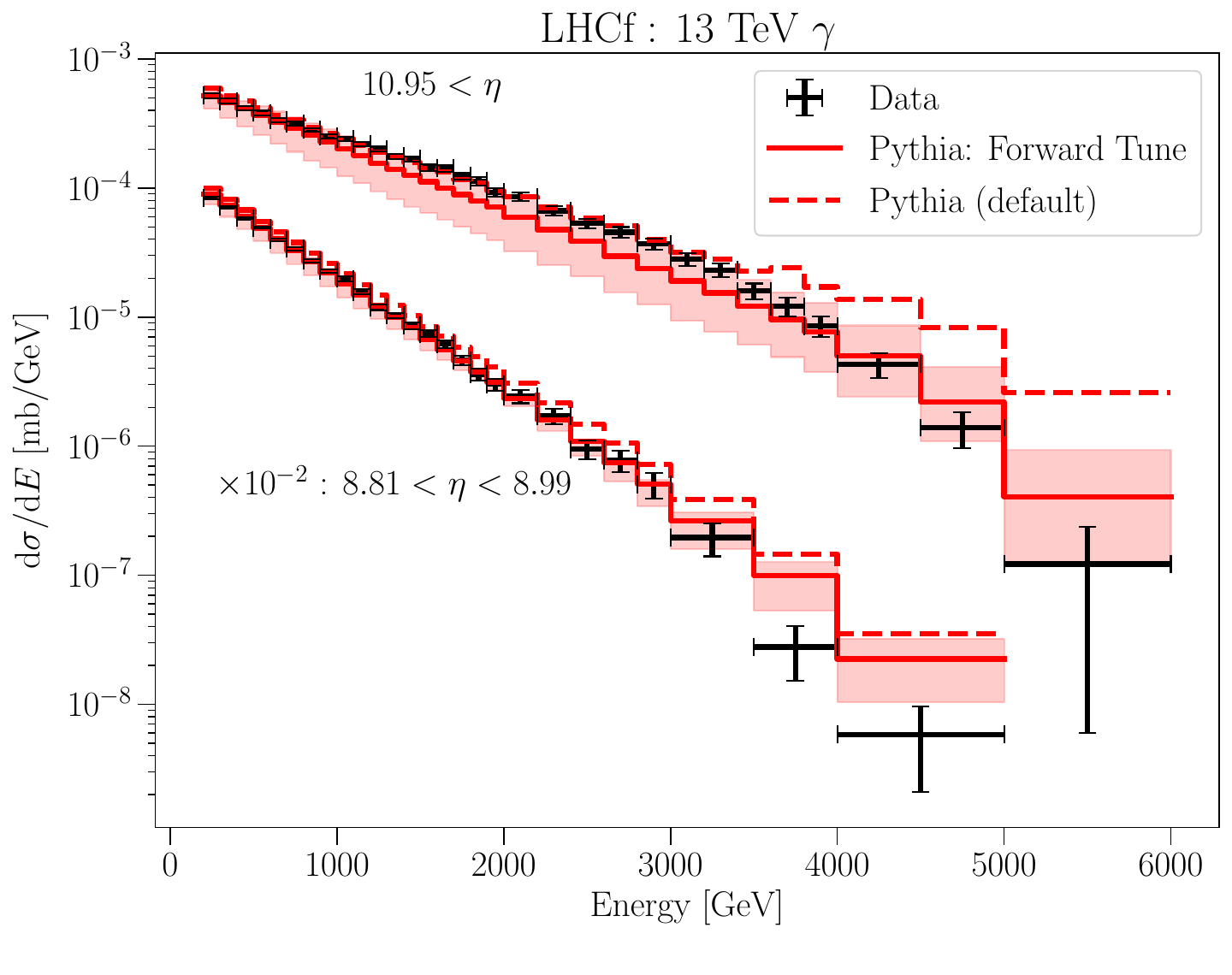}
  \includegraphics[width=0.48\textwidth]{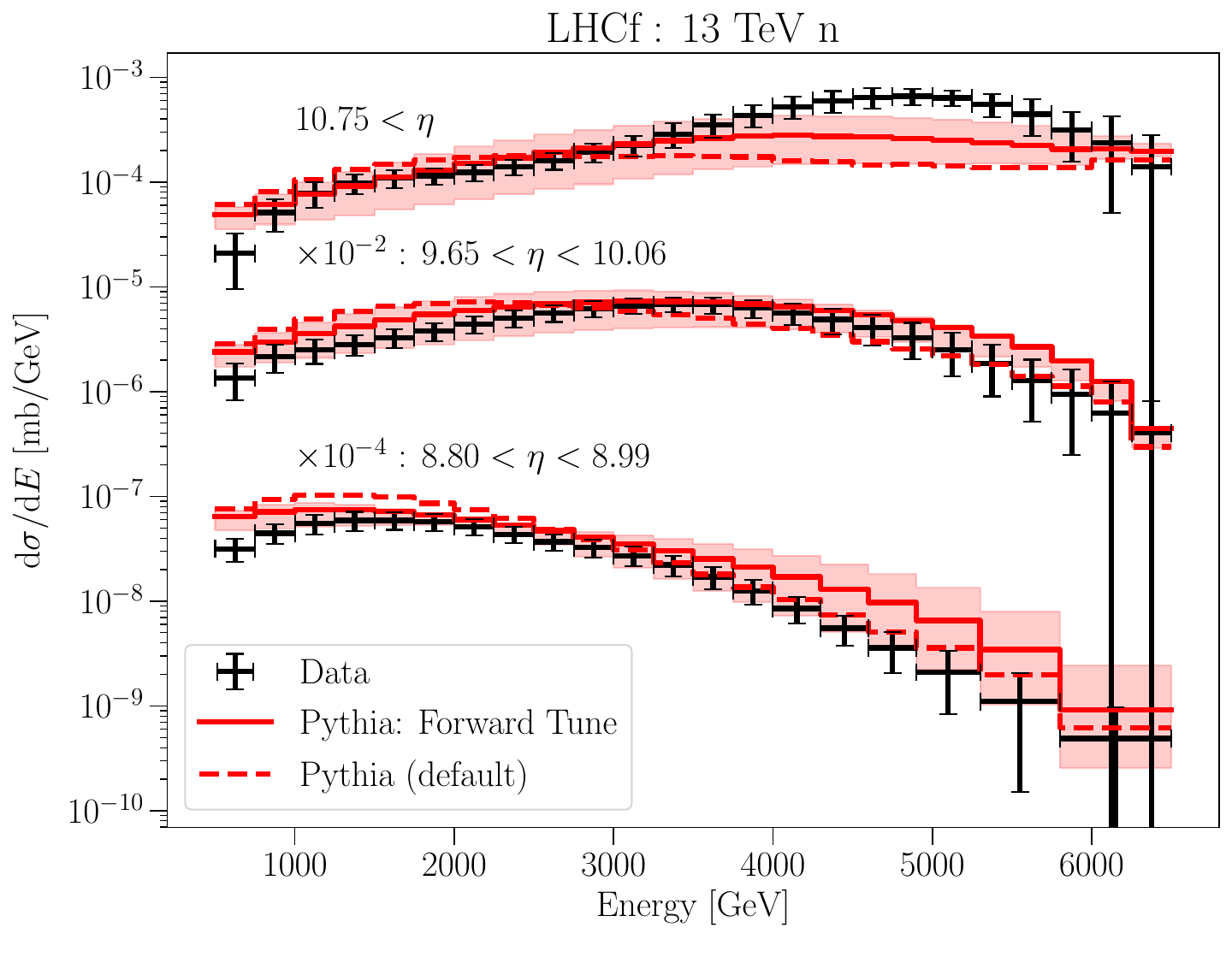}
\caption{LHCf measurements of pions (upper left), photons (lower left) and neutrons at $\sqrt{s}=7~\tev$ (upper right) and $\sqrt{s}=13~\tev$ (lower right) as compared to our tune and the default \pythia prediction. The solid curve is the central prediction of our forward tune, and the shaded region defines our uncertainty band. The dashed curve is the default \pythia prediction and the black error bars are the measured data points. The text near the curves indicates the $\eta$ (or $\pT$) of the curve, as well as a multiplicative shift that we use for display purposes.}
\label{fig:resultsCR}
\end{figure*}

Turning to the tuned LHCf particle spectra, we show our results in \cref{fig:resultsCR}. Here, we show the baseline QCDCR prediction (dashed), our obtained forward physics tune result (solid), and its error band (shaded band) against LHCf measurements.

The pion and photon spectra show similar behavior as most of the photons come from pion decay, so we discuss them together. The pion (photon) spectra can be found in the upper (lower) left panel of \cref{fig:resultsCR}. For the pion spectra, two $\pT$ bins are excluded for display purposes, but this discussion also applies to them. We see that the default configuration predicts too many particles, with a pronounced excess for the most forward bins at high $p_z , E$. Our tune greatly reduces this excess at $E_{\pi^0,\gamma} \approx 3~\tev$ energies, which can in large part be attributed to the removal of the popcorn mechanism on the beam remnant. At smaller momenta, $p_z\sim \tev$, the default curves do better for the largest $\eta$ (smallest $\pT$) pion (photon) bins, but this is small improvement compared to the excess that are reduced in other bins. For most curves, our uncertainty band envelopes most of the data points with the exception of some curves which are still in tension (e.g. pions with $0.8<p_{\perp} {\rm [GeV/c]}<1.0$).

The predicted and measured neutron spectra are shown in the upper (lower) right panels of \cref{fig:resultsCR} for LHC centre-of-mass energies of 7 TeV (13 TeV). For the $\sqrt{s}~=~13~\tev$ neutrons, we show three of six representative $\eta$ bins. The most clear deficiency of the default \pythia prediction is an underproduction of neutrons with $\eta>10.76$, resulting in a spectrum that peaks at lower energies relative to the measured peak. As with the pions and photons, by disabling the popcorn mechanism on the beam remnant our tune can address this deficiency at both LHC energies by producing more hard neutrons. 
\medskip

\begin{figure*}[tbh]
  \includegraphics[width=0.48\textwidth]{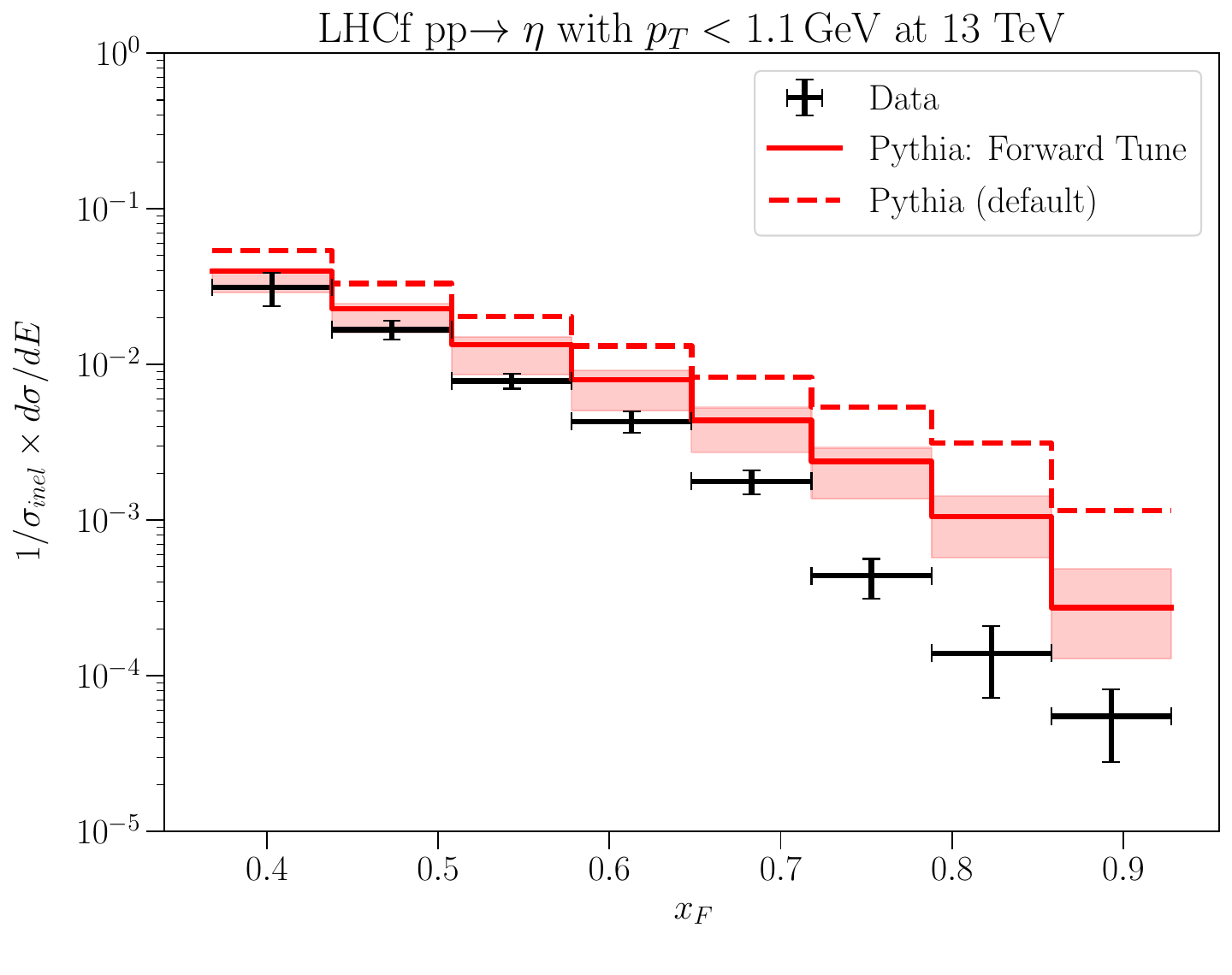}
  \includegraphics[width=0.48\textwidth]{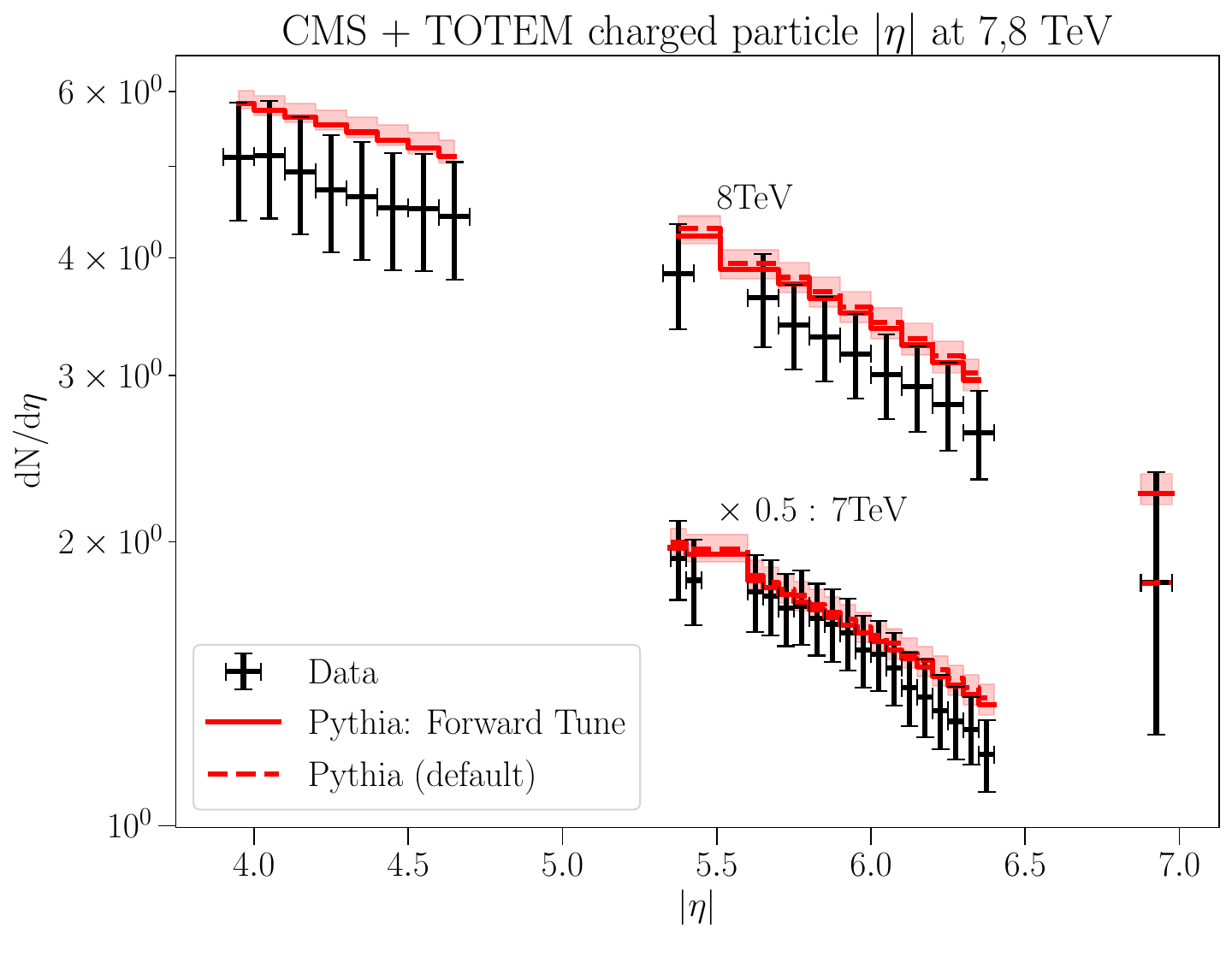}
  \includegraphics[width=0.48\textwidth]{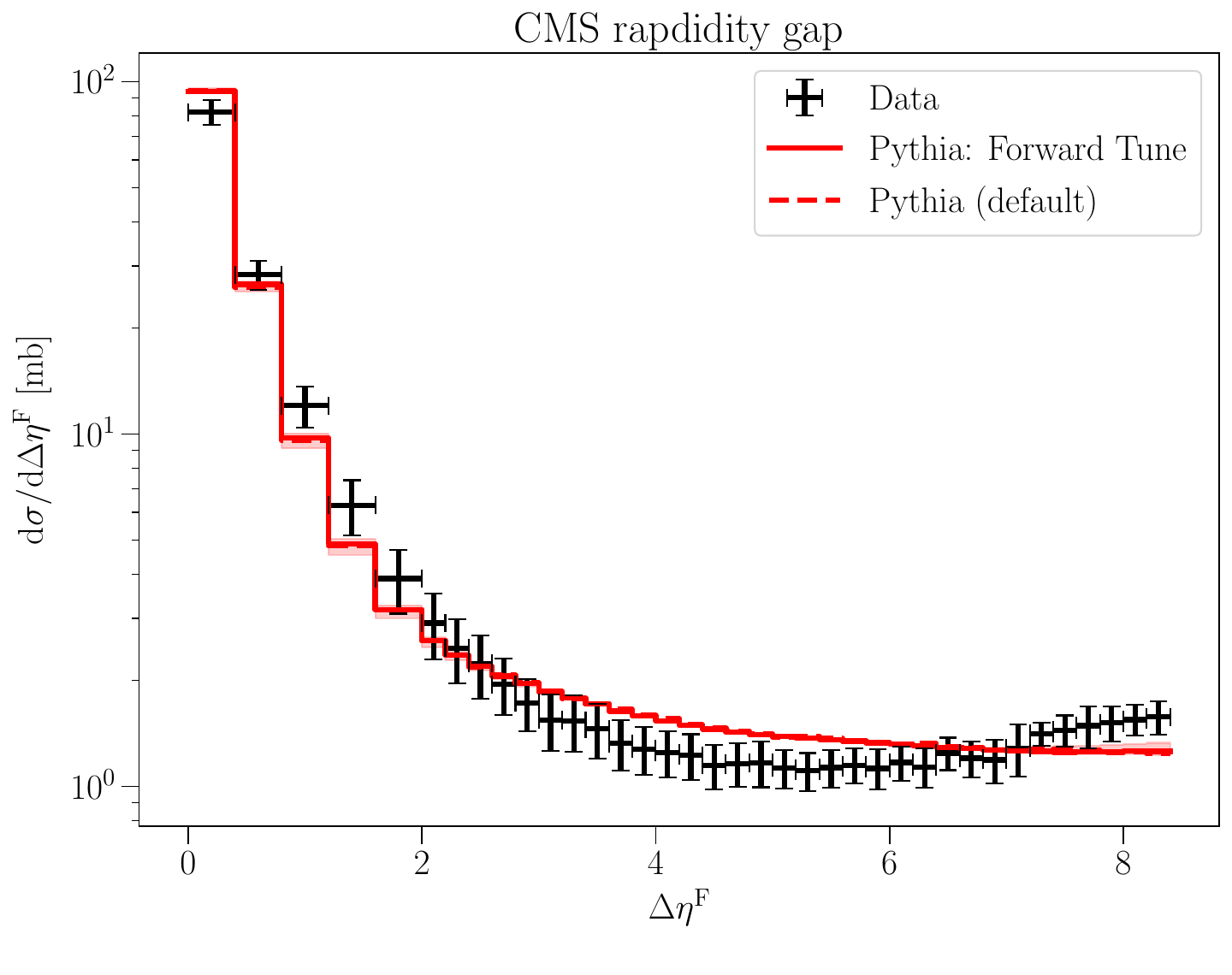}
  \includegraphics[width=0.48\textwidth]{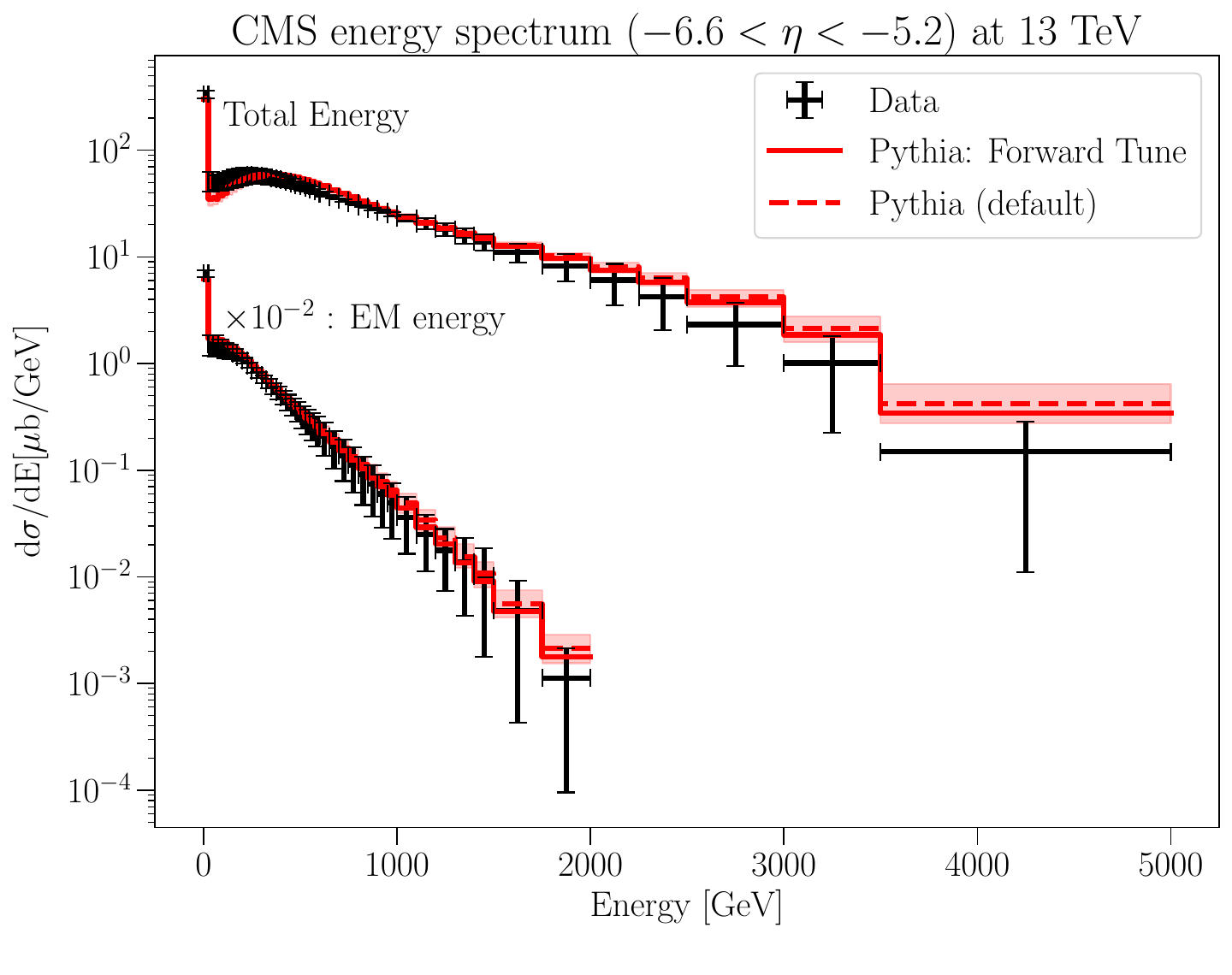}
\caption{In the upper left panel we show the $\eta$ meson distribution as measured by LHCf~\cite{Adriani:2023tyb}. Our tune (solid red) improves on this distribution, as compared to the default configuration (dashed red) ~\cite{Christiansen:2015yqa}. In the remaining panels we compare our tune to more central measurements. In particular we show CMS and TOTEM charged particle pseudorapidity distribution~\cite{CMS:2015inp, CMS:2014kix, TOTEM:2012kvo} (upper right), CMS rapidity gap measurement~\cite{CMS:2015inp} (lower left), and CMS energy spectrum from $-6.6<\eta<-5.2$ ~\cite{CMS:2017dou} (lower right). These measurements are expected to be the msot sensitive to our tuning parameters and we see a small deviation from the default prediction.}
\label{fig:TOTEM_CMS}
\end{figure*}

We show the impact of our tune on the forward $\eta$ meson distribution as measured by LHCf in the upper left panel of \cref{fig:TOTEM_CMS}. The default \pythia configuration overpredicts the number of $\eta$ mesons by almost two orders of magnitude for some bins. While we did not tune to this dataset at all, we see that our tune improves on this by producing less $\eta$'s. 

In the remaining panels of \cref{fig:TOTEM_CMS}, we show our tune as compared to the rapidity distribution of charged particles at CMS and TOTEM's T2 telescope~\cite{CMS:2015inp, CMS:2014kix, TOTEM:2012kvo}, measurements of the rapidity gap distribution at CMS~\cite{CMS:2015inp}, and the energy spectrum measured by CMS' CASTOR calorimeter at $-6.6<\eta<-5.2$~\cite{CMS:2017dou}. There is also a similar rapidity gap analysis from ATLAS~\cite{ATLAS:2012djz} that we checked but do not show, which in addition to the CMS rapidity gap was used to tune the parameters in \pythia associated with the modelling of the elastic, diffractive and total cross section ~\cite{Rasmussen:2018dgo}. 
Besides LHCf, these measurements are the most sensitive to the beam remnant, with TOTEM, and CASTOR covering $\eta \sim 5 \dots 7$ respectively. If our tune had an impact on central physics, we would expect to see an effect on the predicted spectra at these experiments, with a  sub-leading impact on predictions of the rapidity gap at CMS and ATLAS.
In all cases we find a negligible difference between our forward physics tune and the default \pythia prediction, while our uncertainty band produces at most a 5\% variation (seen in the CMS and TOTEM measurements of charged particle pseudorapidity distribution).

\section{Application at FASER}
\label{sec:ApplicationAtFPF}
 
In this section we discuss how our tune can be applied at current and future forward physics experiments. As our tune modifies forward hadron production rates, the decay products of these hadrons will also be affected. Forward hadrons may decay into neutrinos and as a result produce a highly collimated intense neutrino beam along the collision axis of the LHC. Similarly, these hadrons might also decay into yet undiscovered light and weakly interacting particles. As the LHC ring curves away, these neutrinos and BSM particles will travel unimpeded down the collision axis. A new class of experiments has recently begun operating to exploit this particle flux.

One of these experiments is FASER~\cite{FASER:2022hcn}, which is located along the collision axis, 480m downstream of the ATLAS IP, and covers $\eta\gtrsim9$. Located at the front of the experiment is the FASER$\nu$ neutrino detector which is a $25~\cm \times 25~\cm \times 1~\m$ tungsten emulsion detector~\cite{FASER:2019dxq, FASER:2020gpr}. The FASER detector also consists of a long-lived particle detector which searches for the decay products of BSM particles via a series of trackers and a calorimeter. The SND@LHC experiment is also currently taking data, and is located 480m from IP on the opposite side of the ATLAS as FASER~\cite{SNDLHC:2022ihg}. SND@LHC collects off-axis neutrinos from the $pp$ collision, and covers $7.2<\eta<8.7$

To fully utilize the HL-LHC era, upgrades to these experiments have been envisioned, as well as the implementation of further forward physics experiments. These proposed experiments would be located in the FPF~\cite{Anchordoqui:2021ghd, Feng:2022inv}, which is a dedicated cavern for forward physics, located 620~m from the ATLAS IP with space to host a suite of experiments. This includes three detectors aimed at studying neutrinos as well as FASER2 for long-lived particle decays and the FORMOSA experiment for milli-charged particle detection. 

In the following, we apply our tune to make predictions for neutrino fluxes and the dark photon search sensitivity at FASER. These predictions can of course also be applied for other experiments at the FPF. 

\subsection{Neutrinos}

\begin{figure*}[tbh]
  \includegraphics[width=0.99\textwidth]{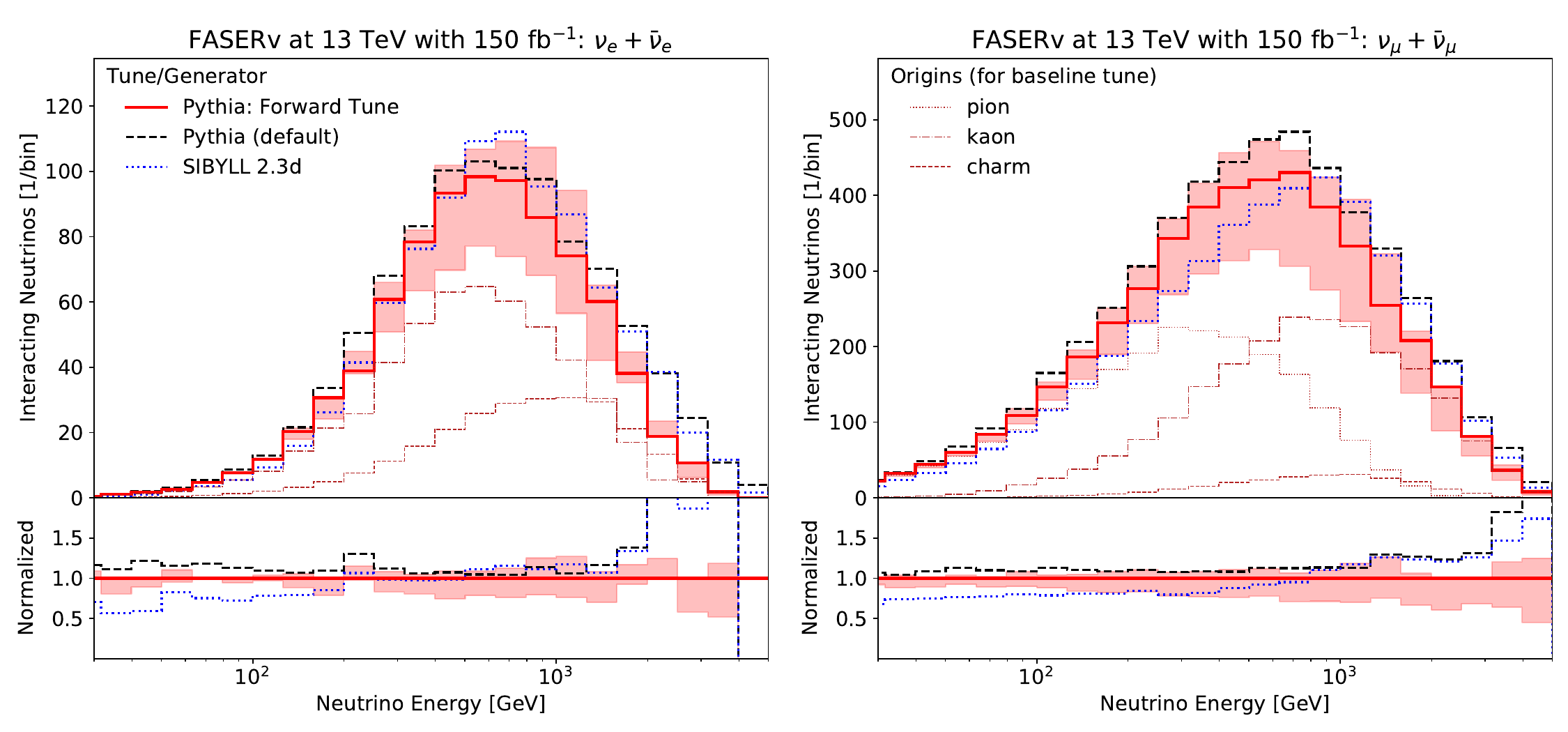}
\caption{Predicted neutrino energy spectrum at FASER$\nu$ for $\nu_e+\bar{\nu}_e$ (left) and $\nu_{\mu}+\bar{\nu}_{\mu}$ (right). The solid red curve is the spectrum computed using the neutrino flux from our tune and the shaded region is our uncertainty band. The dotted, dash-dotted and dashed red curves show the composition of the neutron flux in terms of the parent meson. For comparison we show the interaction spectrum predicted by the default \pythia configuration (dashed black) as well as the \sibyll event generator (dotted blue). In the bottom panel of each figure, we show the ratio of the curves to our tune - our uncertainty analysis gives about a $20\%-30\%$ uncertainty on the interacting neutrino spectrum.}
\label{fig:neutrinos}
\end{figure*}

The LHC produces an intense flux of high energy neutrinos. This has been first realized in the 1980's~\cite{DeRujula:1992sn} but no detection has been made until recently. The first candidates were first detected using the FASER$\nu$ pilot detector in 2021~\cite{FASER:2021mtu}, and further observed by the FASER detector in 2023~\cite{FASER:2023zcr}. These neutrinos are expected to originate from pion, kaon, and charm meson decays.

The first estimate of the neutrino flux was provided in Ref.~\cite{Kling:2021gos}, which takes into account both the prompt flux from charm meson decay occurring at the IP, and the displaced decays of long-lived hadrons. This estimate uses a variety of MC event generators from cosmic-ray physics (\textsc{Epos-Lhc}~\cite{Pierog:2013ria} , \textsc{Sibyll~2.3d}~\cite{Riehn:2019jet}, \textsc{QgsJet~2.04}~\cite{Ostapchenko:2010vb}, \textsc{DpmJet~3.2019.1}~\cite{Fedynitch:2015kcn}) as well as \pythia to model the hadron production at the LHC. The average and spread of these generators have then been used to define a first rough neutrino flux estimate and its uncertainty.

Using our improved forward physics tune, we make predictions for the event rate at FASER$\nu$. For this, we use the dedicated fast simulation as introduced in Ref.~\cite{Kling:2021gos} to model the production and decay of long-lived hadrons when passing through the LHC beam pipe and magnetic fields. We have updated the magnet field configuration to those used at the beginning of Run 3, and use the same beam crossing angle of 160 $\mu$rad downwards. We then convolute the neutrino flux obtained using \pythia with the interaction cross-sections obtained from \textsc{Genie}~\cite{Andreopoulos:2009rq} to calculate the number of expected events in FASER$\nu$.

Our results are shown in \cref{fig:neutrinos} for an integrated luminosity of 150~fb$^{-1}$. The left and right panel are the electron and muon neutrino spectrum, respectively. The red line is our central prediction for our forward tune, and the dashed black line is the spectrum with the default configuration of \pythia. The red shaded region is our uncertainty band as determined in \cref{sec:uncertainty}. For comparison we also show the predictions from the \textsc{Sibyll} event generator. In the bottom panel we show the ratios of the curves to our tuned curve - we see that our uncertainty gives roughly a 20\% uncertainty in the neutrino interaction rate.

Also indicated in \cref{fig:neutrinos} is composition of the neutrinos in terms of their parent mesons, shown in dotted, dash-dotted, and dashed curves for pion, kaon, and charm meson respectively. Clearly, the majority of electron neutrinos come from kaon decay, with a significant charm component at higher energies. Muon neutrinos on the other hand, are dominantly produced by pion decay at lower energies, and kaon decay at high energies. While \pythia models charm production, we note that there are ongoing efforts to provide refined predictions of forward charm production mode using perturbative QCD \cite{Bai:2020ukz, Maciula:2022lzk, Bhattacharya:2023zei, BKRSpaper}, some of which predict significantly enhanced charm production rates. In the regime where light hadron decays dominate the neutrino composition, the obtained flux uncertainty with our tune roughly agrees with that of Ref.~\cite{Kling:2021gos} 

We note that currently, we only include uncertainties associated with the kinematic distribution. There could be additional sources of uncertainties associated with the flavor composition, especially the kaon to pion production fraction. Indeed, observation from astroparticle physics suggest that forward kaon production might be different than predicted by existing hadronic interaction models. Over more than two decades, cosmic ray experiments have reported significant discrepancies between the number of observed muons in high-energy cosmic ray air showers and model predictions~\cite{PierreAuger:2014ucz, PierreAuger:2016nfk, Soldin:2021wyv}. This observation is commonly referred to as the muon puzzle. Extensive studies have suggested that an enhanced rate of strangeness production in the forward direction could explain the discrepancy~\cite{Allen:2013hfa, Anchordoqui:2016oxy, Albrecht:2021cxw}. While forward strange measurements could shed light on this discrepancy, no attempt was made to include this in our tune due to the lack of data. 

\subsection{Dark Photons}

\begin{figure*}[tbh]
  \includegraphics[width=0.48\textwidth]{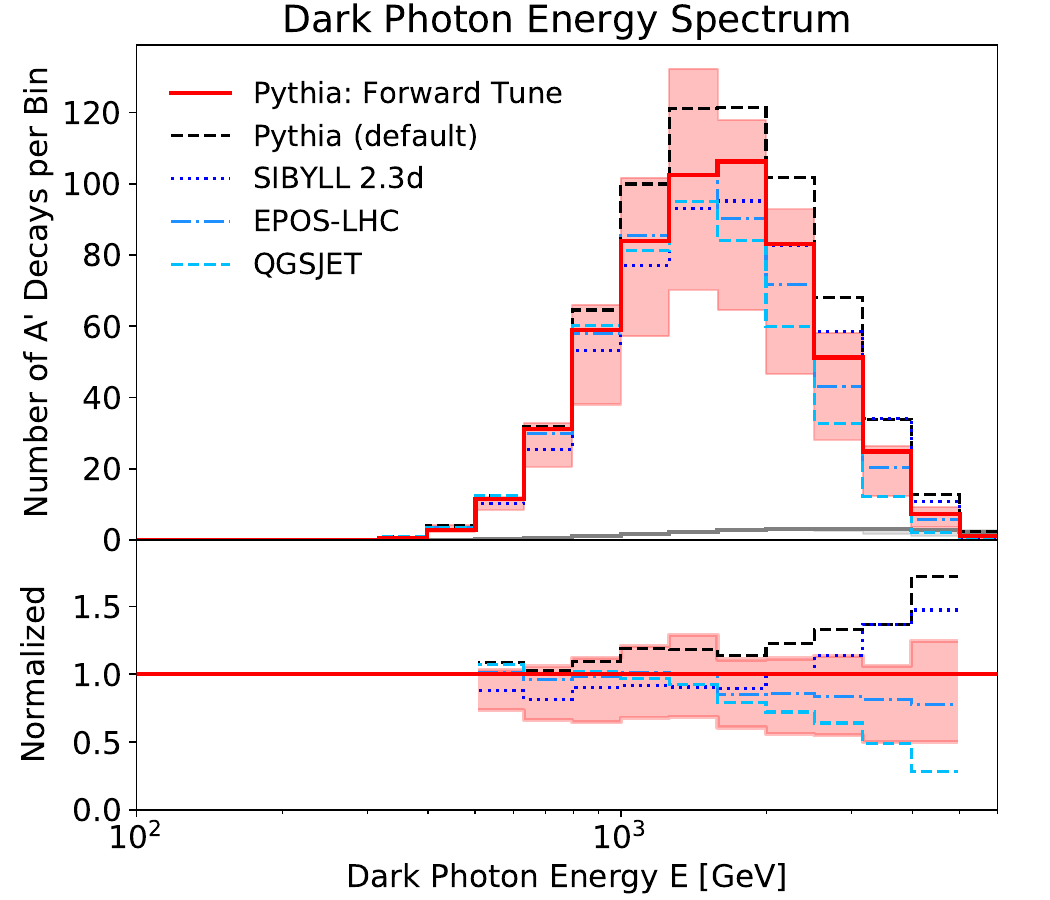}
  \includegraphics[width=0.48\textwidth]{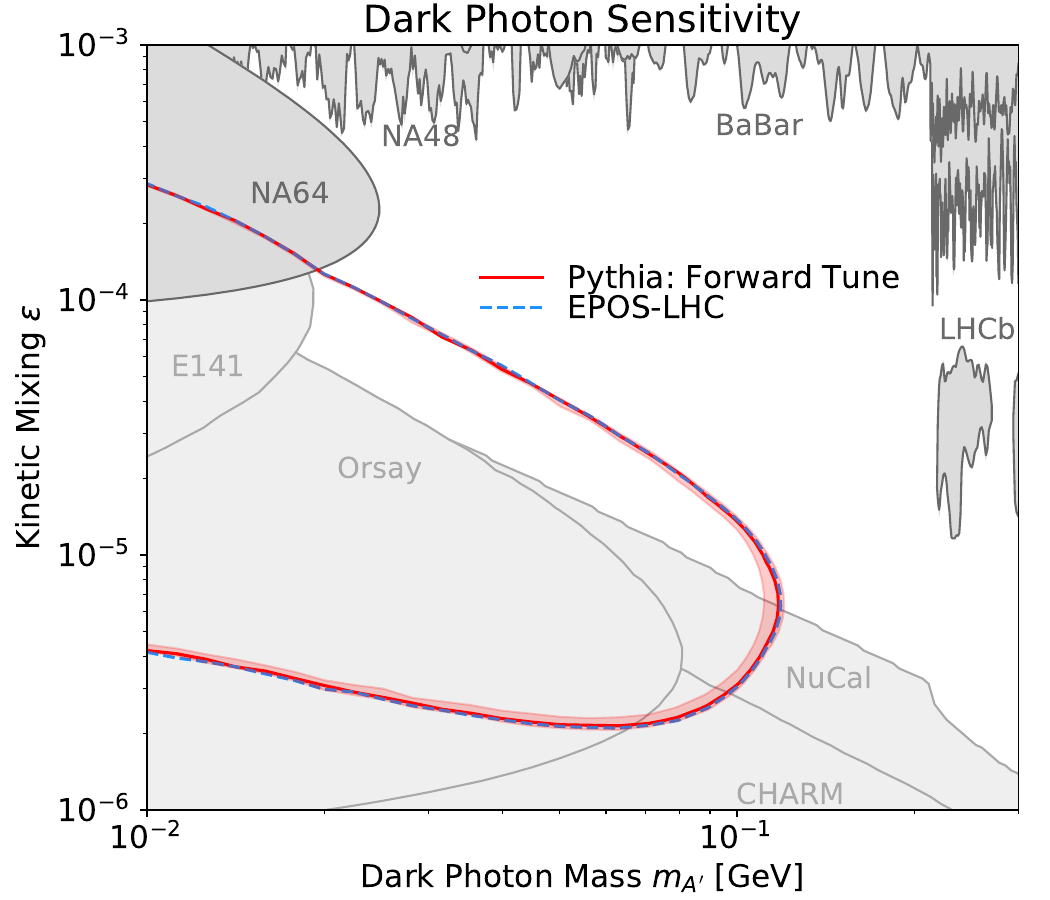}
\caption{Dark photon spectrum at FASER for $m_{A'},\epsilon = 25~{\rm MeV} , 3\times10^{-5}$ (left) and the discovery reach for FASER using the spectrum predicted by our tune. In the left panel, we show the spectra predicted by our tune (solid red) as well as the associated uncertainty that we calculate (shaded red). For comparison we show the spectra predicted by the default \pythia configuration (dashed black), \sibyll (dotted blue), \eposlhc (dash-dotted blue) and \qgsjet (dashed blue). In the bottom section of the left panel, we show the ratio of the curves to our tune and see that our uncertainty imparts a $\approx 50\%$ uncertainty on the number of dark photon decays in FASER. In the right panel we show FASER's sensitivity in dark photon parameter space for our tune (solid red), our associated uncertainty (shaded red) and the sensitivity predicted by \eposlhc for comparison (dashed blue). While the uncertainty we calculate has a small impact on FASER's sensitivity, see that the uncertainty is most important when FASER is limited by exposure (i.e. at small $\epsilon$, large $m_{A'}$).}
\label{fig:darkphoton}
\end{figure*}

The other main purpose of FASER is the search for light long-lived particles with MeV-GeV masses~\cite{Feng:2017uoz, FASER:2018ceo, FASER:2018eoc}. These are for example motivated by dark matter and, more generally, dark sectors. One of the primary examples discussed in the literature is dark photons. The dark photon is a gauge field of a broken U(1) in the dark sector. Through kinetic mixing with the SM photon, the dark photon, $A'$, can interact with SM fields. This interaction is suppressed by the kinetic mixing parameter $\epsilon$ with an interaction Lagrangian, ${\cal L}\supset \epsilon/2~ F'^{\mu\nu}F_{\mu \nu}$ where $F$ ($F'$) is the field strength of the (dark) photon. For massive dark photons with $2m_e < m_{A'} < 2m_\mu$, the dark photon will primarily decay into $e^+e^-$. With sufficiently small $\epsilon$, the dark photon will travel several hundred meters before decaying and could decay inside FASER which has been designed to detect this signal.

Recently, FASER reported the first results for the search for dark photons~\cite{FASER:2023tle}. In the probed regime, dark photons mainly come from neutral pion decay with small contributions from eta-meson decay and dark bremsstrahlung. The FASER collaboration has estimated the dark photon flux using \textsc{Epos-Lhc}. The signal uncertainty was estimated by comparing with \textsc{Sibyll} and \textsc{QgsJet}. 

We use our \pythia forward physics tune to model forward particle production and \textsc{Foresee}~\cite{Kling:2021fwx} to then obtain the expected dark photon event rate at FASER. The left panel of \cref{fig:darkphoton} shows the energy spectrum of dark photons decaying in FASER during Run3 with 150 $\rm{fb}^{-1}$ integrated luminosity for $m_{A'} = 25$ MeV and $\epsilon = 3\times 10^{-5}$. This point lies at the edge of the previously excluded region. The red curve is our main prediction, and the shaded band is error band. The bottom panel shows the ratio of the curves to our central prediction and shows that our uncertainty is roughly 30\%. For comparison, we also show the dark photon spectrum from the default \pythia configuration (dashed black) and the prediction from \textsc{Sibyll}, \textsc{Epos-Lhc}, and \textsc{QgsJet} in dotted, dash-dotted, dashed blue curves. We can see that the predictions from these other generators are consistent with our prediction. We note that our uncertainty is slightly larger than the uncertainty obtained by comparing generators at low energy and similar at higher energy. 

The right panel shows the FASER sensitivity for Run~3 with 150~fb$^{-1}$ in the dark photon parameter space spanned by $\epsilon$ and $m_{A'}$. The gray shaded areas are excluded by existing experiments (from above by prompt resonance searches, from below by long-lived particle searches in beam dumps) as obtained from \textsc{DarkCast}~\cite{Ilten:2018crw, Baruch:2022esd}. The constraints shown in light gray are obtained by recasting experimental analyses while dark gray bounds were reported directly by the experiments. Using our tune we draw our expected sensitivity contour in red with our uncertainty as the shaded contour, and compare with  sensitivity contour as calculated with \eposlhc in dashed blue. We find that the sensitivity calculated with each configuration is comparable. We also note the overall effect of the flux uncertainty on the sensitivity reach is small. This is due to an exponential ($\epsilon^4$) suppression of the event rate at large (small) $\epsilon$. 

\section{Conclusion}
\label{sec:Conclusion}

In recent years, a new set of experiments has begun their operation in the forward direction of the LHC, with the purpose of observing and studying collider neutrinos as well as searching for light long-lived particles. This emerging forward neutrino and particle search program requires precise predictions of the anticipated particle fluxes. Currently, forward particle production is often simulated using specialized MC event generators developed for cosmic ray physics, such as \eposlhcfs, \qgsjetfs and \sibyllfs. Additionally, multipurpose event generators like \pythia can also be utilized. However, it has been noticed that the corresponding predicted spectra exhibit some discrepancies when compared to the measured flux obtained from the LHCf experiment.

This paper addresses this issue by introducing a new dedicated forward tune for \pythiafs, specifically designed for forward physics studies at the LHC. This newly proposed tune is based on the QCDCR scenario introduced in Ref.~\cite{Christiansen:2015yqa}, and offers a more adaptable approach for modeling beam remnant hadronization and is tuned to match the available forward particle spectra measured by LHCf. A comprehensive list of the relevant parameters and their corresponding values can be found in \cref{table:parms}. We also explored an alternative tune based on the well-established Monash configuration utilizing the default CR scenario. However, we found that this alternative tune exhibits a poorer agreement with LHCf data compared to the QCDCR-based approach, as discussed in \cref{sec:monashdiscussion}.

When fine-tuning event generators, the process currently lacks a well-established method for quantifying and incorporating measures of uncertainty. In addition to our fit, we also provide an uncertainty in a data-driven way for the first time. What has sometimes been done, is to take the spread in event generators' predictions to define an uncertainty band on the true particle distribution. In this paper, we vary the relevant tuning parameter, $\sigma$, around the best-fit such that 68\% of the data points are captured. This band can then be used for further applications to study the impact of flux uncertainties.

To demonstrate an application of our tune, we also show its impact on the predicted neutrino and dark photon sensitivity. A precise understanding of the neutrino flux that better agrees with forward physics data is important to study TeV neutrino interactions, and an improved understanding of the dark photon flux will increase experiments' search sensitivity. Our tune also provides a means of understanding the flux uncertainty in each case. For both cases, we find that our tune is consistent with the \sibyllfs, \eposlhcfs and \qgsjetfs generators, and that our uncertainty band is a bit wider than the spread of these generators' predictions.

In conclusion, our forward tune of \pythia enables enhanced precision in the exploration of forward physics phenomena. Our approach presents a data-guided mechanism for honing the neutrino flux and its associated uncertainty. By gaining better control over the uncertainty in neutrino flux, it opens the gateway to improved investigations, including a refined modeling of neutrino production through hadron decay~\cite{fluxpaper}, exploration of sterile neutrino production, and a deeper understanding of neutrino interactions within experiments designed to unveil proton structure~\cite{Fieg:2023ss}, and potential avenues toward uncovering new signatures of physics.

\section*{Acknowledgment}

We thank Aki Ariga, Tomoko Ariga, Eugenio Berti, Andy Buckley, Anatoli Fedynitch, Jonathan Feng, Hiroaki Menjo, Kenneth Österberg, Stefan Hoeche, Tanguy Pierog, Christophe Royon for many fruitful discussions. We are grateful to the authors and maintainers of many open-source software packages, including \texttt{scikit-hep}~\cite{Rodrigues:2020syo}. This work utilized the infrastructure for high-performance and high-throughput computing, research data storage and analysis, and scientific software tool integration built, operated, and updated by the Research Cyberinfrastructure Center (RCIC) at the University of California, Irvine (UCI). The RCIC provides cluster-based systems, application software, and scalable storage to directly support the UCI research community. \url{https://rcic.uci.edu}. The work of M.F. was supported by NSF Grant PHY-2210283 and was also supported by NSF Graduate Research Fellowship Award No. DGE-1839285. F.K. acknowledges support by the Deutsche Forschungsgemeinschaft under Germany's Excellence Strategy - EXC 2121 Quantum Universe - 390833306. T.S. has been supported by the Swedish Research Council, contract number 2016-05996.

\appendix
\vspace{-2mm}
\section{Alternate Monash Tune}
\label{sec:monashdiscussion}

\begin{figure*}[tbh]
  \includegraphics[width=0.48\textwidth]{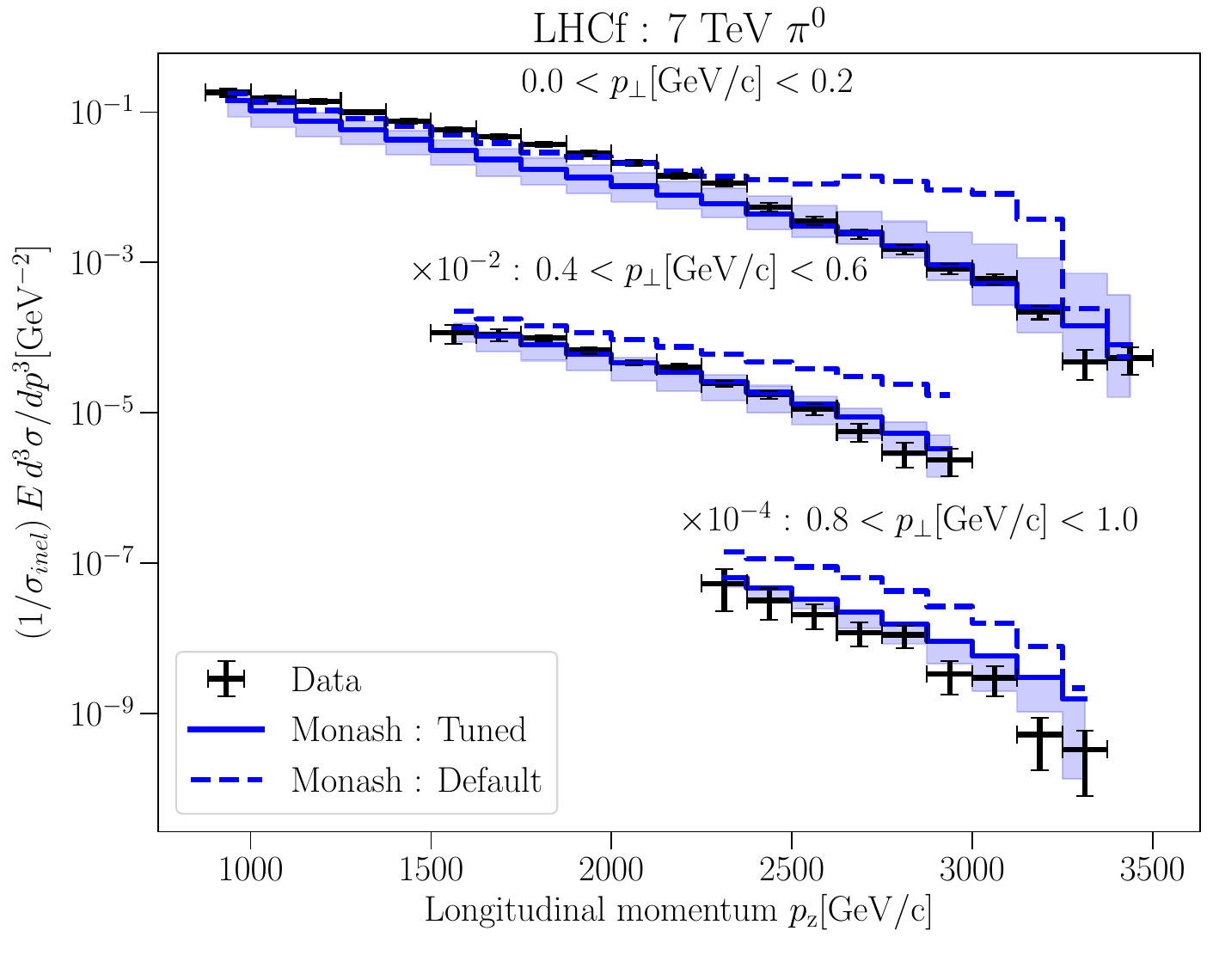}
  \includegraphics[width=0.48\textwidth]{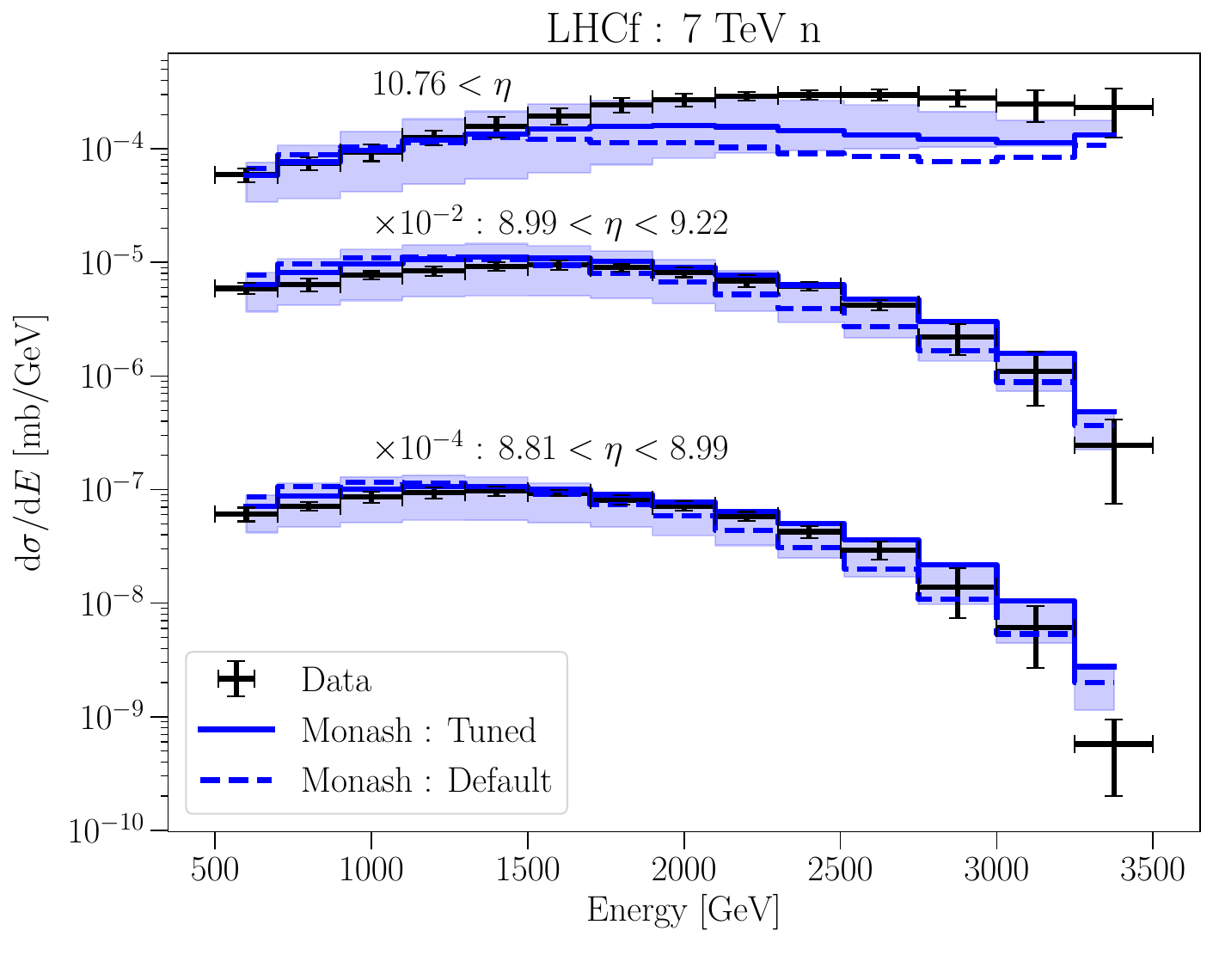}
  \includegraphics[width=0.48\textwidth]{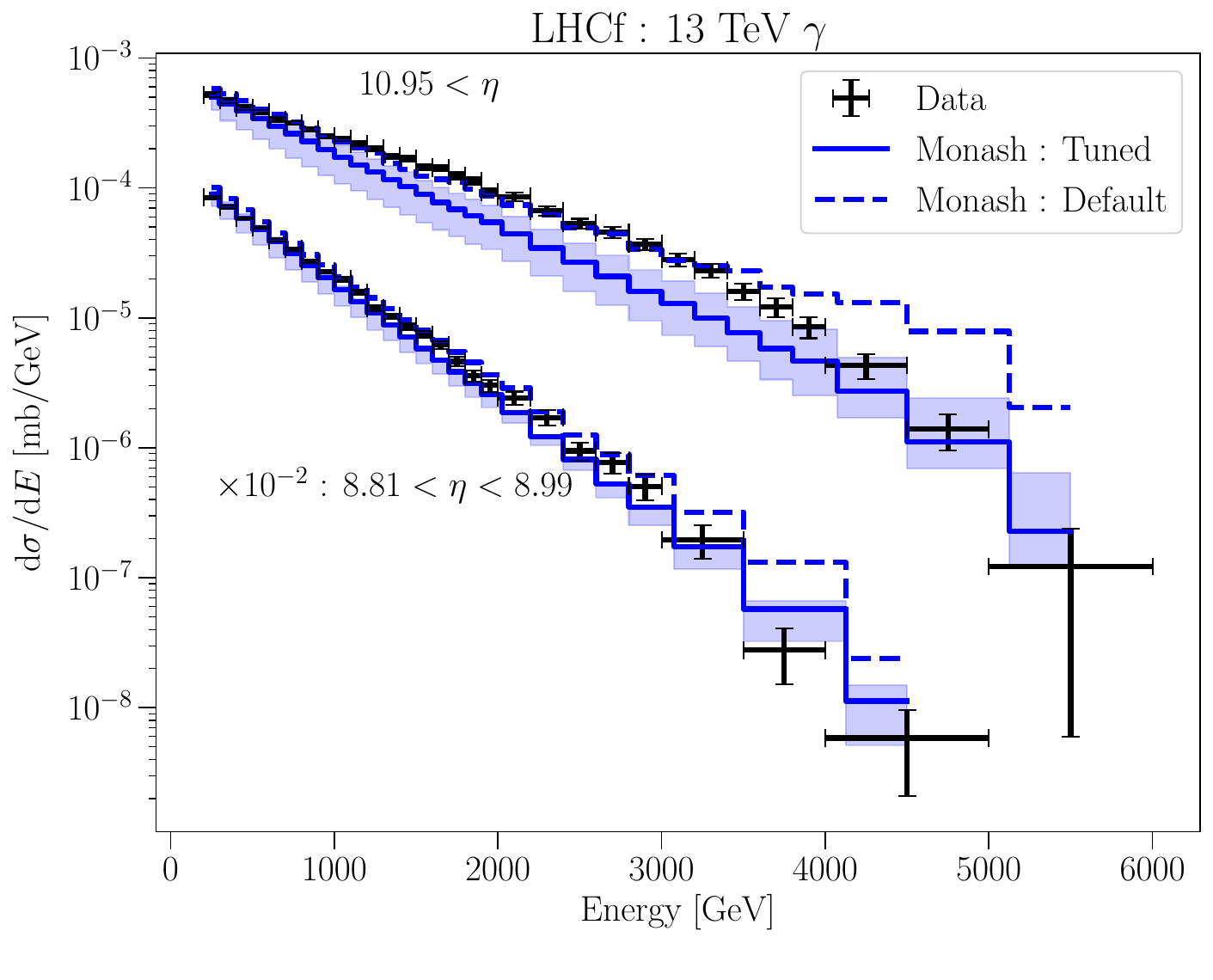}
  \includegraphics[width=0.48\textwidth]{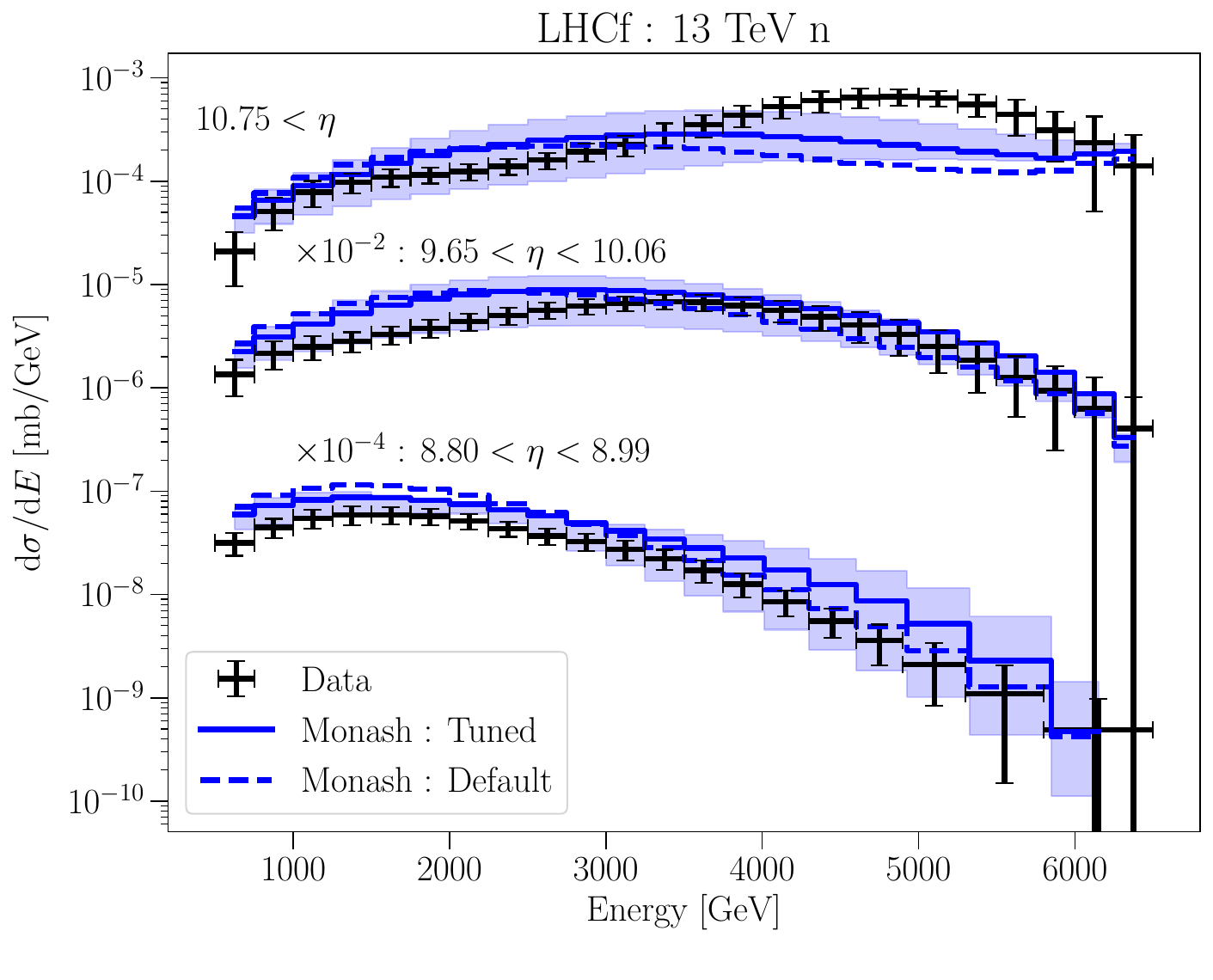}
\caption{LHCf spectra as compared to an alternate tune that we explored based on the Monash tune. The LHCf measurements of pions (upper left), photons (lower left) and neutrons at $\sqrt{s}=7~\tev$ (upper right) and $\sqrt{s}=13~\tev$ (lower right) as compared to our alternate, Monash based tune and the default \pythia prediction. The solid curve is the central prediction of our Monash based tune, and the shaded region defines our uncertainty band. The dashed curve is the default \pythia prediction and the black error bars are the measured data points. The text near the curves indicates the $\eta$ (or $\pT$) of the curve, as well as a multiplicative shift that we use for display purposes.}
\label{fig:resultsMonash}
\end{figure*}

\begin{table*}[t]
\begin{tabular}{l|c||c|c|c}
\hline\hline
 Full name & Shorthand & Baseline (Monash) & Forward Tune & Uncertainty\\ \hline
 \texttt{BeamRemnants:dampPopcorn} & $d_{\mathrm{pop}}$ & 1 & 0 &\\ 
 \texttt{BeamRemnants:hardRemnantBaryon} & $f_{\mathrm{remn}}$ & off & on & \\
 \texttt{BeamRemnants:aRemnantBaryon} & $a_{\mathrm{remn}}$ & - & 0.68 & \\
 \texttt{BeamRemnants:bRemnantBaryon} & $b_{\mathrm{remn}}$ & - & 1.22 &\\
 \texttt{BeamRemnants:primordialKTsoft} & $\sigma_{\mathrm{soft}}$ & 0.9 & 0.56 & $0.2 \dots 1.42$ \\
 \texttt{BeamRemnants:primordialKThard} & $\sigma_{\mathrm{hard}}$ & 1.8 & 1.8& \\
 \texttt{BeamRemnants:halfScaleForKT} & $Q_{\mathrm{half}}$ & 1.5 & 10 &\\
 \texttt{BeamRemnants:halfMassForKT} & $m_{\mathrm{half}}$ & 1 & 1 &\\
 \texttt{BeamRemnants:primordialKTremnant} & $\sigma_{\mathrm{remn}}$ & 0.4 & 0.56 & $0.22 \dots 1.42$ \\
 \hline \hline
\end{tabular}
\caption{The main \pythia parameters studied in this article, their default parameters in the Monash tune, and their values in the Monash based tune obtained in this study. The last column shows the uncertainty range for $\sigma_{\mathrm{soft}} = \sigma_{\mathrm{remn}}$ as discussed in \cref{sec:uncertainty}.}
\label{table:alternatefits}
\end{table*}

Here we discuss and show the results of the alternate tune which is based off the well-known Monash tune to central physics, which we provide for comparison purposes. We show our fitting results in \cref{table:alternatefits} and our fitted spectra against LHCf data in \cref{fig:resultsMonash}. While we find comparable tuning parameters for this Monash based tune as our main tune, the QCDCR configuration from Ref.~\cite{Christiansen:2015yqa} proves to be an important feature for our tuning purposes. 

While the Monash based tune has some same advantages of our primary tune, there are some clear deficiencies. In particular, the photon spectra shows a significant underproduction of forward photons with $E\lesssim 3~\tev$ - a similar effect same can be seen for relatively softer pions. A further deficiency can be seen in the $\eta>10.76$ neutron spectra, particularly for the $\sqrt{s} = 7~\tev$ --- the Monash tune does not address the shape of the neutron spectra as well as our primary tune does.


\bibliography{references}

\providecommand{\href}[2]{#2}\begingroup\raggedright\begin{thebibliography}{10}

\bibitem{FASER:2023tle}
{\bf FASER} Collaboration, H.~Abreu {\em et al.}, ``{Search for Dark Photons
  with the FASER detector at the LHC},''
  \href{http://arxiv.org/abs/2308.05587}{{\tt arXiv:2308.05587 [hep-ex]}}.

\bibitem{FASER:2021mtu}
{\bf FASER} Collaboration, H.~Abreu {\em et al.}, ``{First neutrino interaction
  candidates at the LHC},''
  \href{http://dx.doi.org/10.1103/PhysRevD.104.L091101}{{\em Phys. Rev. D} {\bf
  104} (2021) no.~9, L091101}, \href{http://arxiv.org/abs/2105.06197}{{\tt
  arXiv:2105.06197 [hep-ex]}}.

\bibitem{FASER:2023zcr}
{\bf FASER} Collaboration, H.~Abreu {\em et al.}, ``{First Direct Observation
  of Collider Neutrinos with FASER at the LHC},''
  \href{http://arxiv.org/abs/2303.14185}{{\tt arXiv:2303.14185 [hep-ex]}}.

\bibitem{Anchordoqui:2021ghd}
L.~A. Anchordoqui {\em et al.}, ``{The Forward Physics Facility: Sites,
  experiments, and physics potential},''
  \href{http://dx.doi.org/10.1016/j.physrep.2022.04.004}{{\em Phys. Rept.} {\bf
  968} (2022)  1--50}, \href{http://arxiv.org/abs/2109.10905}{{\tt
  arXiv:2109.10905 [hep-ph]}}.

\bibitem{Feng:2022inv}
J.~L. Feng {\em et al.}, ``{The Forward Physics Facility at the High-Luminosity
  LHC},'' \href{http://dx.doi.org/10.1088/1361-6471/ac865e}{{\em J. Phys. G}
  {\bf 50} (2023) no.~3, 030501}, \href{http://arxiv.org/abs/2203.05090}{{\tt
  arXiv:2203.05090 [hep-ex]}}.

\bibitem{Sjostrand:2014zea}
T.~Sj\"ostrand, S.~Ask, J.~R. Christiansen, R.~Corke, N.~Desai, P.~Ilten,
  S.~Mrenna, S.~Prestel, C.~O. Rasmussen, and P.~Z. Skands, ``{An introduction
  to PYTHIA 8.2},'' \href{http://dx.doi.org/10.1016/j.cpc.2015.01.024}{{\em
  Comput. Phys. Commun.} {\bf 191} (2015)  159--177},
  \href{http://arxiv.org/abs/1410.3012}{{\tt arXiv:1410.3012 [hep-ph]}}.

\bibitem{Bierlich:2022pfr}
C.~Bierlich {\em et al.}, ``{A comprehensive guide to the physics and usage of
  PYTHIA 8.3},'' \href{http://arxiv.org/abs/2203.11601}{{\tt arXiv:2203.11601
  [hep-ph]}}.

\bibitem{Skands:2014pea}
P.~Skands, S.~Carrazza, and J.~Rojo, ``{Tuning PYTHIA 8.1: the Monash 2013
  Tune},'' \href{http://dx.doi.org/10.1140/epjc/s10052-014-3024-y}{{\em Eur.
  Phys. J. C} {\bf 74} (2014) no.~8, 3024},
  \href{http://arxiv.org/abs/1404.5630}{{\tt arXiv:1404.5630 [hep-ph]}}.

\bibitem{LHCf:2015nel}
{\bf LHCf} Collaboration, O.~Adriani {\em et al.}, ``{Measurement of very
  forward neutron energy spectra for 7 TeV proton\textendash{}proton collisions
  at the Large Hadron Collider},''
  \href{http://dx.doi.org/10.1016/j.physletb.2015.09.041}{{\em Phys. Lett. B}
  {\bf 750} (2015)  360--366}, \href{http://arxiv.org/abs/1503.03505}{{\tt
  arXiv:1503.03505 [hep-ex]}}.

\bibitem{LHCf:2020hjf}
{\bf LHCf} Collaboration, O.~Adriani {\em et al.}, ``{Measurement of energy
  flow, cross section and average inelasticity of forward neutrons produced in
  $ \sqrt{s} $ = 13 TeV proton-proton collisions with the LHCf Arm2
  detector},'' \href{http://dx.doi.org/10.1007/JHEP07(2020)016}{{\em JHEP} {\bf
  07} (2020)  016}, \href{http://arxiv.org/abs/2003.02192}{{\tt
  arXiv:2003.02192 [hep-ex]}}.

\bibitem{LHCf:2015rcj}
{\bf LHCf} Collaboration, O.~Adriani {\em et al.}, ``{Measurements of
  longitudinal and transverse momentum distributions for neutral pions in the
  forward-rapidity region with the LHCf detector},''
  \href{http://dx.doi.org/10.1103/PhysRevD.94.032007}{{\em Phys. Rev. D} {\bf
  94} (2016) no.~3, 032007}, \href{http://arxiv.org/abs/1507.08764}{{\tt
  arXiv:1507.08764 [hep-ex]}}.

\bibitem{LHCf:2017fnw}
{\bf LHCf} Collaboration, O.~Adriani {\em et al.}, ``{Measurement of forward
  photon production cross-section in proton\textendash{}proton collisions at
  $\sqrt{s}$ = 13 TeV with the LHCf detector},''
  \href{http://dx.doi.org/10.1016/j.physletb.2017.12.050}{{\em Phys. Lett. B}
  {\bf 780} (2018)  233--239}, \href{http://arxiv.org/abs/1703.07678}{{\tt
  arXiv:1703.07678 [hep-ex]}}.

\bibitem{Feynman:1969ej}
R.~P. Feynman, ``{Very high-energy collisions of hadrons},''
  \href{http://dx.doi.org/10.1103/PhysRevLett.23.1415}{{\em Phys. Rev. Lett.}
  {\bf 23} (1969)  1415--1417}.

\bibitem{Andersson:1983ia}
B.~Andersson, G.~Gustafson, G.~Ingelman, and T.~Sj{\"o}strand, ``{Parton
  Fragmentation and String Dynamics},''
  \href{http://dx.doi.org/10.1016/0370-1573(83)90080-7}{{\em Phys. Rept.} {\bf
  97} (1983)  31--145}.

\bibitem{Sjostrand:1987su}
T.~Sj{\"o}strand and M.~van Zijl, ``{A Multiple Interaction Model for the Event
  Structure in Hadron Collisions},''
  \href{http://dx.doi.org/10.1103/PhysRevD.36.2019}{{\em Phys. Rev. D} {\bf 36}
  (1987)  2019}.

\bibitem{Christiansen:2015yqa}
J.~R. Christiansen and P.~Z. Skands, ``{String Formation Beyond Leading
  Colour},'' \href{http://dx.doi.org/10.1007/JHEP08(2015)003}{{\em JHEP} {\bf
  08} (2015)  003}, \href{http://arxiv.org/abs/1505.01681}{{\tt
  arXiv:1505.01681 [hep-ph]}}.

\bibitem{CMS:2019uws}
{\bf CMS} Collaboration, A.~M. Sirunyan {\em et al.}, ``{Production of
  $\Lambda_\mathrm{c}^+$ baryons in proton-proton and lead-lead collisions at
  $\sqrt{s_\mathrm{NN}}=$ 5.02 TeV},''
  \href{http://dx.doi.org/10.1016/j.physletb.2020.135328}{{\em Phys. Lett. B}
  {\bf 803} (2020)  135328}, \href{http://arxiv.org/abs/1906.03322}{{\tt
  arXiv:1906.03322 [hep-ex]}}.

\bibitem{ALICE:2020wla}
{\bf ALICE} Collaboration, S.~Acharya {\em et al.}, ``{$\Lambda^+_c$ production
  in $pp$ and in $p$-Pb collisions at $\sqrt {s_{NN}}$=5.02 TeV},''
  \href{http://dx.doi.org/10.1103/PhysRevC.104.054905}{{\em Phys. Rev. C} {\bf
  104} (2021) no.~5, 054905}, \href{http://arxiv.org/abs/2011.06079}{{\tt
  arXiv:2011.06079 [nucl-ex]}}.

\bibitem{Ingelman:1984ns}
G.~Ingelman and P.~E. Schlein, ``{Jet Structure in High Mass Diffractive
  Scattering},'' \href{http://dx.doi.org/10.1016/0370-2693(85)91181-5}{{\em
  Phys. Lett. B} {\bf 152} (1985)  256--260}.

\bibitem{LHCf:2006kzv}
{\bf LHCf} Collaboration, O.~Adriani {\em et al.}, ``{Technical design report
  of the LHCf experiment: Measurement of photons and neutral pions in the very
  forward region of LHC},''.

\bibitem{Adriani:2023tyb}
O.~Adriani {\em et al.}, ``{Measurement of the forward $\eta$ meson production
  rate in p-p collisions at $\sqrt{s}$=13 TeV with the LHCf-Arm2 detector},''
  \href{http://arxiv.org/abs/2305.06633}{{\tt arXiv:2305.06633 [hep-ex]}}.

\bibitem{Krishnamoorthy:2021nwv}
M.~Krishnamoorthy, H.~Schulz, X.~Ju, W.~Wang, S.~Leyffer, Z.~Marshall,
  S.~Mrenna, J.~M\"uller, and J.~B. Kowalkowski, ``{Apprentice for Event
  Generator Tuning},''
  \href{http://dx.doi.org/10.1051/epjconf/202125103060}{{\em EPJ Web Conf.}
  {\bf 251} (2021)  03060}, \href{http://arxiv.org/abs/2103.05748}{{\tt
  arXiv:2103.05748 [hep-ex]}}.

\bibitem{CMS:2015inp}
{\bf CMS} Collaboration, V.~Khachatryan {\em et al.}, ``{Measurement of
  diffraction dissociation cross sections in pp collisions at $\sqrt{s}$ = 7
  TeV},'' \href{http://dx.doi.org/10.1103/PhysRevD.92.012003}{{\em Phys. Rev.
  D} {\bf 92} (2015) no.~1, 012003},
  \href{http://arxiv.org/abs/1503.08689}{{\tt arXiv:1503.08689 [hep-ex]}}.

\bibitem{CMS:2014kix}
{\bf CMS, TOTEM} Collaboration, S.~Chatrchyan {\em et al.}, ``{Measurement of
  pseudorapidity distributions of charged particles in proton-proton collisions
  at $\sqrt{s}$ = 8 TeV by the CMS and TOTEM experiments},''
  \href{http://dx.doi.org/10.1140/epjc/s10052-014-3053-6}{{\em Eur. Phys. J. C}
  {\bf 74} (2014) no.~10, 3053}, \href{http://arxiv.org/abs/1405.0722}{{\tt
  arXiv:1405.0722 [hep-ex]}}.

\bibitem{TOTEM:2012kvo}
{\bf TOTEM} Collaboration, G.~Antchev {\em et al.}, ``{Measurement of the
  forward charged particle pseudorapidity density in $pp$ collisions at
  $\sqrt{s} = 7$ TeV with the TOTEM experiment},''
  \href{http://dx.doi.org/10.1209/0295-5075/98/31002}{{\em EPL} {\bf 98} (2012)
  no.~3, 31002}, \href{http://arxiv.org/abs/1205.4105}{{\tt arXiv:1205.4105
  [hep-ex]}}.

\bibitem{CMS:2017dou}
{\bf CMS} Collaboration, A.~M. Sirunyan {\em et al.}, ``{Measurement of the
  inclusive energy spectrum in the very forward direction in proton-proton
  collisions at $ \sqrt{s}=13 $ TeV},''
  \href{http://dx.doi.org/10.1007/JHEP08(2017)046}{{\em JHEP} {\bf 08} (2017)
  046}, \href{http://arxiv.org/abs/1701.08695}{{\tt arXiv:1701.08695
  [hep-ex]}}.

\bibitem{ATLAS:2012djz}
{\bf ATLAS} Collaboration, G.~Aad {\em et al.}, ``{Rapidity gap cross sections
  measured with the ATLAS detector in $pp$ collisions at $\sqrt{s}=7$ TeV},''
  \href{http://dx.doi.org/10.1140/epjc/s10052-012-1926-0}{{\em Eur. Phys. J. C}
  {\bf 72} (2012)  1926}, \href{http://arxiv.org/abs/1201.2808}{{\tt
  arXiv:1201.2808 [hep-ex]}}.

\bibitem{Rasmussen:2018dgo}
C.~O. Rasmussen and T.~Sj\"ostrand, ``{Models for total, elastic and
  diffractive cross sections},''
  \href{http://dx.doi.org/10.1140/epjc/s10052-018-5940-8}{{\em Eur. Phys. J. C}
  {\bf 78} (2018) no.~6, 461}, \href{http://arxiv.org/abs/1804.10373}{{\tt
  arXiv:1804.10373 [hep-ph]}}.

\bibitem{FASER:2022hcn}
{\bf FASER} Collaboration, H.~Abreu {\em et al.}, ``{The FASER Detector},''
  \href{http://arxiv.org/abs/2207.11427}{{\tt arXiv:2207.11427
  [physics.ins-det]}}.

\bibitem{FASER:2019dxq}
{\bf FASER} Collaboration, H.~Abreu {\em et al.}, ``{Detecting and Studying
  High-Energy Collider Neutrinos with FASER at the LHC},''
  \href{http://dx.doi.org/10.1140/epjc/s10052-020-7631-5}{{\em Eur. Phys. J. C}
  {\bf 80} (2020) no.~1, 61}, \href{http://arxiv.org/abs/1908.02310}{{\tt
  arXiv:1908.02310 [hep-ex]}}.

\bibitem{FASER:2020gpr}
{\bf FASER} Collaboration, H.~Abreu {\em et al.}, ``{Technical Proposal:
  FASERnu},'' \href{http://arxiv.org/abs/2001.03073}{{\tt arXiv:2001.03073
  [physics.ins-det]}}.

\bibitem{SNDLHC:2022ihg}
{\bf SND@LHC} Collaboration, G.~Acampora {\em et al.}, ``{SND@LHC: The
  Scattering and Neutrino Detector at the LHC},''
  \href{http://arxiv.org/abs/2210.02784}{{\tt arXiv:2210.02784 [hep-ex]}}.

\bibitem{DeRujula:1992sn}
A.~De~Rujula, E.~Fernandez, and J.~J. Gomez-Cadenas, ``{Neutrino fluxes at
  future hadron colliders},''
  \href{http://dx.doi.org/10.1016/0550-3213(93)90427-Q}{{\em Nucl. Phys. B}
  {\bf 405} (1993)  80--108}.

\bibitem{Kling:2021gos}
F.~Kling and L.~J. Nevay, ``{Forward neutrino fluxes at the LHC},''
  \href{http://dx.doi.org/10.1103/PhysRevD.104.113008}{{\em Phys. Rev. D} {\bf
  104} (2021) no.~11, 113008}, \href{http://arxiv.org/abs/2105.08270}{{\tt
  arXiv:2105.08270 [hep-ph]}}.

\bibitem{Pierog:2013ria}
T.~Pierog, I.~Karpenko, J.~M. Katzy, E.~Yatsenko, and K.~Werner, ``{EPOS LHC:
  Test of collective hadronization with data measured at the CERN Large Hadron
  Collider},'' \href{http://dx.doi.org/10.1103/PhysRevC.92.034906}{{\em Phys.
  Rev. C} {\bf 92} (2015) no.~3, 034906},
  \href{http://arxiv.org/abs/1306.0121}{{\tt arXiv:1306.0121 [hep-ph]}}.

\bibitem{Riehn:2019jet}
F.~Riehn, R.~Engel, A.~Fedynitch, T.~K. Gaisser, and T.~Stanev, ``{Hadronic
  interaction model Sibyll 2.3d and extensive air showers},''
  \href{http://dx.doi.org/10.1103/PhysRevD.102.063002}{{\em Phys. Rev. D} {\bf
  102} (2020) no.~6, 063002}, \href{http://arxiv.org/abs/1912.03300}{{\tt
  arXiv:1912.03300 [hep-ph]}}.

\bibitem{Ostapchenko:2010vb}
S.~Ostapchenko, ``{Monte Carlo treatment of hadronic interactions in enhanced
  Pomeron scheme: I. QGSJET-II model},''
  \href{http://dx.doi.org/10.1103/PhysRevD.83.014018}{{\em Phys. Rev. D} {\bf
  83} (2011)  014018}, \href{http://arxiv.org/abs/1010.1869}{{\tt
  arXiv:1010.1869 [hep-ph]}}.

\bibitem{Fedynitch:2015kcn}
A.~Fedynitch, \href{http://dx.doi.org/10.5445/IR/1000055433}{{\em {Cascade
  equations and hadronic interactions at very high energies}}}.
\newblock PhD thesis, KIT, Karlsruhe, Dept. Phys., 11, 2015.

\bibitem{Andreopoulos:2009rq}
C.~Andreopoulos {\em et al.}, ``{The GENIE Neutrino Monte Carlo Generator},''
  \href{http://dx.doi.org/10.1016/j.nima.2009.12.009}{{\em Nucl. Instrum. Meth.
  A} {\bf 614} (2010)  87--104}, \href{http://arxiv.org/abs/0905.2517}{{\tt
  arXiv:0905.2517 [hep-ph]}}.

\bibitem{Bai:2020ukz}
W.~Bai, M.~Diwan, M.~V. Garzelli, Y.~S. Jeong, and M.~H. Reno, ``{Far-forward
  neutrinos at the Large Hadron Collider},''
  \href{http://dx.doi.org/10.1007/JHEP06(2020)032}{{\em JHEP} {\bf 06} (2020)
  032}, \href{http://arxiv.org/abs/2002.03012}{{\tt arXiv:2002.03012
  [hep-ph]}}.

\bibitem{Maciula:2022lzk}
R.~Maciula and A.~Szczurek, ``{Far-forward production of charm mesons and
  neutrinos at forward physics facilities at the LHC and the intrinsic charm in
  the proton},'' \href{http://dx.doi.org/10.1103/PhysRevD.107.034002}{{\em
  Phys. Rev. D} {\bf 107} (2023) no.~3, 034002},
  \href{http://arxiv.org/abs/2210.08890}{{\tt arXiv:2210.08890 [hep-ph]}}.

\bibitem{Bhattacharya:2023zei}
A.~Bhattacharya, F.~Kling, I.~Sarcevic, and A.~M. Stasto, ``{Forward Neutrinos
  from Charm at Large Hadron Collider},''
  \href{http://arxiv.org/abs/2306.01578}{{\tt arXiv:2306.01578 [hep-ph]}}.

\bibitem{BKRSpaper}
L.~Buonocore, F.~Kling, L.~Rottoli, and J.~Sominka, ``Predictions for neutrinos
  and new physics from forward heavy hadron production at the lhc (in
  prep.),''.

\bibitem{PierreAuger:2014ucz}
{\bf Pierre Auger} Collaboration, A.~Aab {\em et al.}, ``{Muons in Air Showers
  at the Pierre Auger Observatory: Mean Number in Highly Inclined Events},''
  \href{http://dx.doi.org/10.1103/PhysRevD.91.032003}{{\em Phys. Rev. D} {\bf
  91} (2015) no.~3, 032003}, \href{http://arxiv.org/abs/1408.1421}{{\tt
  arXiv:1408.1421 [astro-ph.HE]}}. [Erratum: Phys.Rev.D 91, 059901 (2015)].

\bibitem{PierreAuger:2016nfk}
{\bf Pierre Auger} Collaboration, A.~Aab {\em et al.}, ``{Testing Hadronic
  Interactions at Ultrahigh Energies with Air Showers Measured by the Pierre
  Auger Observatory},''
  \href{http://dx.doi.org/10.1103/PhysRevLett.117.192001}{{\em Phys. Rev.
  Lett.} {\bf 117} (2016) no.~19, 192001},
  \href{http://arxiv.org/abs/1610.08509}{{\tt arXiv:1610.08509 [hep-ex]}}.

\bibitem{Soldin:2021wyv}
{\bf EAS-MSU, IceCube, KASCADE-Grande, NEVOD-DECOR, Pierre Auger, SUGAR,
  Telescope Array, Yakutsk EAS Array} Collaboration, D.~Soldin, ``{Update on
  the Combined Analysis of Muon Measurements from Nine Air Shower
  Experiments},'' \href{http://dx.doi.org/10.22323/1.395.0349}{{\em PoS} {\bf
  ICRC2021} (2021)  349}, \href{http://arxiv.org/abs/2108.08341}{{\tt
  arXiv:2108.08341 [astro-ph.HE]}}.

\bibitem{Allen:2013hfa}
J.~Allen and G.~Farrar, ``{Testing models of new physics with UHE air shower
  observations},'' in {\em {33rd International Cosmic Ray Conference}},
  p.~1182.
\newblock 7, 2013.
\newblock \href{http://arxiv.org/abs/1307.7131}{{\tt arXiv:1307.7131
  [astro-ph.HE]}}.

\bibitem{Anchordoqui:2016oxy}
L.~A. Anchordoqui, H.~Goldberg, and T.~J. Weiler, ``{Strange fireball as an
  explanation of the muon excess in Auger data},''
  \href{http://dx.doi.org/10.1103/PhysRevD.95.063005}{{\em Phys. Rev. D} {\bf
  95} (2017) no.~6, 063005}, \href{http://arxiv.org/abs/1612.07328}{{\tt
  arXiv:1612.07328 [hep-ph]}}.

\bibitem{Albrecht:2021cxw}
J.~Albrecht {\em et al.}, ``{The Muon Puzzle in cosmic-ray induced air showers
  and its connection to the Large Hadron Collider},''
  \href{http://dx.doi.org/10.1007/s10509-022-04054-5}{{\em Astrophys. Space
  Sci.} {\bf 367} (2022) no.~3, 27},
  \href{http://arxiv.org/abs/2105.06148}{{\tt arXiv:2105.06148 [astro-ph.HE]}}.

\bibitem{Feng:2017uoz}
J.~L. Feng, I.~Galon, F.~Kling, and S.~Trojanowski, ``{ForwArd Search
  ExpeRiment at the LHC},''
  \href{http://dx.doi.org/10.1103/PhysRevD.97.035001}{{\em Phys. Rev. D} {\bf
  97} (2018) no.~3, 035001}, \href{http://arxiv.org/abs/1708.09389}{{\tt
  arXiv:1708.09389 [hep-ph]}}.

\bibitem{FASER:2018ceo}
{\bf FASER} Collaboration, A.~Ariga {\em et al.}, ``{Letter of Intent for
  FASER: ForwArd Search ExpeRiment at the LHC},''
  \href{http://arxiv.org/abs/1811.10243}{{\tt arXiv:1811.10243
  [physics.ins-det]}}.

\bibitem{FASER:2018eoc}
{\bf FASER} Collaboration, A.~Ariga {\em et al.}, ``{FASER\textquoteright{}s
  physics reach for long-lived particles},''
  \href{http://dx.doi.org/10.1103/PhysRevD.99.095011}{{\em Phys. Rev. D} {\bf
  99} (2019) no.~9, 095011}, \href{http://arxiv.org/abs/1811.12522}{{\tt
  arXiv:1811.12522 [hep-ph]}}.

\bibitem{Kling:2021fwx}
F.~Kling and S.~Trojanowski, ``{Forward experiment sensitivity estimator for
  the LHC and future hadron colliders},''
  \href{http://dx.doi.org/10.1103/PhysRevD.104.035012}{{\em Phys. Rev. D} {\bf
  104} (2021) no.~3, 035012}, \href{http://arxiv.org/abs/2105.07077}{{\tt
  arXiv:2105.07077 [hep-ph]}}.

\bibitem{Ilten:2018crw}
P.~Ilten, Y.~Soreq, M.~Williams, and W.~Xue, ``{Serendipity in dark photon
  searches},'' \href{http://dx.doi.org/10.1007/JHEP06(2018)004}{{\em JHEP} {\bf
  06} (2018)  004}, \href{http://arxiv.org/abs/1801.04847}{{\tt
  arXiv:1801.04847 [hep-ph]}}.

\bibitem{Baruch:2022esd}
C.~Baruch, P.~Ilten, Y.~Soreq, and M.~Williams, ``{Axial vectors in
  DarkCast},'' \href{http://dx.doi.org/10.1007/JHEP11(2022)124}{{\em JHEP} {\bf
  11} (2022)  124}, \href{http://arxiv.org/abs/2206.08563}{{\tt
  arXiv:2206.08563 [hep-ph]}}.

\bibitem{fluxpaper}
F.~Kling, T.~M{\"a}kel{\"a}, and S.~Trojanowski, ``Investigating the fluxes and
  physics potential of lhc neutrino experiments (in prep.),''.

\bibitem{Fieg:2023ss}
M.~Fieg {\em et al.}, ``The lhc as a neutrino ion collider (in prep.),''.

\bibitem{Rodrigues:2020syo}
E.~Rodrigues {\em et al.}, ``{The Scikit HEP Project - overview and
  prospects},'' \href{http://dx.doi.org/10.1051/epjconf/202024506028}{{\em EPJ
  Web Conf.} {\bf 245} (2020)  06028},
  \href{http://arxiv.org/abs/2007.03577}{{\tt arXiv:2007.03577
  [physics.comp-ph]}}.

\end{thebibliography}\endgroup

\end{document}